# Graph Mining for Cybersecurity: A Survey

BO YAN, Beijing University of Posts and Telecommunications, China
CHENG YANG, Beijing University of Posts and Telecommunications, China
CHUAN SHI*, Beijing University of Posts and Telecommunications, China
YONG FANG, Sichuan University, China
QI LI, Beijing University of Posts and Telecommunications, China
YANFANG YE, University of Notre Dame, USA
JUNPING DU, Beijing University of Posts and Telecommunications, China

The explosive growth of cyber attacks nowadays, such as malware, spam, and intrusions, caused severe consequences on society. Securing cyberspace has become an utmost concern for organizations and governments. Traditional Machine Learning (ML) based methods are extensively used in detecting cyber threats, but they hardly model the correlations between real-world cyber entities. In recent years, with the proliferation of graph mining techniques, many researchers investigated these techniques for capturing correlations between cyber entities and achieving high performance. It is imperative to summarize existing graph-based cybersecurity solutions to provide a guide for future studies. Therefore, as a key contribution of this paper, we provide a comprehensive review of graph mining for cybersecurity, including an overview of cybersecurity tasks, the typical graph mining techniques, and the general process of applying them to cybersecurity, as well as various solutions for different cybersecurity tasks. For each task, we probe into relevant methods and highlight the graph types, graph approaches, and task levels in their modeling. Furthermore, we collect open datasets and toolkits for graph-based cybersecurity. Finally, we outlook the potential directions of this field for future research.

## 1 INTRODUCTION

With the development of the Internet, various cyber attacks occur constantly, not only bringing dramatic losses to individuals and enterprises but also posing a serious threat to the country. For example, it was pointed out by [91] that phishing attacks increased by 600% in March 2020. Another example is that the exposure of Hillary Clinton's email by hackers caused a huge political impact and even interfered with the U.S. election. Cybersecurity

---

*Chuan Shi is a corresponding author. E-mail: shichuan@bupt.edu.cn

Authors' addresses: Bo Yan, boyan@bupt.edu.cn, Beijing University of Posts and Telecommunications, Beijing, China, 100876; Cheng Yang, yangcheng@bupt.edu.cn, Beijing University of Posts and Telecommunications, Beijing, China, 100876; Chuan Shi, shichuan@bupt.edu.cn, Beijing University of Posts and Telecommunications, Beijing, China, 100876; Yong Fang, yfang@scu.edu.cn, Sichuan University, Sichuan, Chengdu, China, 610207; Qi Li, liqi2001@bupt.edu.cn, Beijing University of Posts and Telecommunications, Beijing, China, 100876; Yanfang Ye, yye7@nd.edu, University of Notre Dame, South Bend, Indiana, USA, 46556; Junping Du, junpingdu@126.com, Beijing University of Posts and Telecommunications, Beijing, China, 100876.

has become a key factor affecting global risks, thus drawing widespread concerns from academia and industry nowadays.

To combat cyber attacks, ML-based methods have been widely used in cybersecurity applications. Generally, cybersecurity issues can be modeled as ML tasks. Earlier ML-based solutions mostly use manually constructed application-specific features as inputs, and then train ML models (e.g., Support Vector Machine (SVM) and K-Nearest Neighbor (KNN)) to detect cybersecurity threats. For instance, the Application Programming Interface (API) call sequence is often utilized to construct behavior features, so as to identify the homology of malware [130, 217]. However, earlier ML-based methods heavily rely on feature engineering, which is time-consuming and also limits their generalizability. As a key component of ML, Deep Learning (DL) alleviated these limitations by the automatic learning of high-level cyber attack features, as well as the considerably higher capacity of learning, and thus become desirable approaches in the last ten years [114, 137].

Despite the success of these ML-based methods for cybersecurity applications, there exist many explicit or implicit correlations between real-world cyber entities, like API call relations in Android apps (short for applications), which can characterize the structural patterns of cyber criminals. Unfortunately, traditional ML-based methods hardly capture these correlations, which dramatically hinders performance on some tasks. In recent years, with the surge of study on graph mining methods [54, 60, 86], more and more researchers began to apply graph mining technologies to cybersecurity. Generally, graph mining technologies are good at mining the semantic information and spatial correlations of cyber entities for better cyber attack detection. For example, malware tends to have high density and strong closeness API call graph [7]; Fraudsters are inclined to have similar interaction behaviors which can be modeled by meta-path based graph mining methods [69, 240]. At present, graph mining technologies have been proven to the state-of-the-art in many cybersecurity tasks (e.g., defaulter detection [240] and rumor detection [10]). In addition, many methods have been integrated into the enterprise's security products as a core component [69, 70, 101].

Though many graph mining techniques have been widely used in solving cybersecurity tasks, to the best of our knowledge, there is no comprehensive survey on graph-based cybersecurity applications. However, this kind of survey is urgently needed, considering the increasingly severe environment of cybersecurity. It can provide an overall reference for quickly designing graph-based cybersecurity solutions and also help later researchers avoid repetitive work. We also notice that there are several surveys on ML/DL in cybersecurity [104, 114, 219], as well as some surveys on graph mining techniques in other fields [21, 116, 173, 216]. [74] presents a most relevant survey to ours. It only summarizes the earlier graph mining solutions for capturing propagation patterns of malware. In contrast, our survey covers a wide range of existing graph-based solutions for various cybersecurity tasks.

In this survey, we provide a comprehensive review of graph mining techniques used in cybersecurity. We give a detailed taxonomy and descriptions of various cybersecurity tasks, as well as their applied graph mining techniques. We also formulate the general steps of designing graph-based cybersecurity applications for guidance. After that, detailed graph-based cybersecurity solutions are introduced. Concretely, we classify cybersecurity tasks mainly based on the applied graph mining techniques, to highlight the unique advantages of graph methods in solving current hotspot security issues. Besides, for each cybersecurity task, we present typical graph types and graph mining techniques to guide the development of novel cybersecurity solutions and facilitate the contrast towards existing ones. Moreover, we first collect the typical graph-based public datasets and toolkits used in cybersecurity to facilitate baseline experiments. Through systematically analyzing existing studies, we elaborate keynotes in designing graph-based cybersecurity solutions, as well as some promising directions.

The remainder of this survey is organized as follows. In Section 2, we give a taxonomy and descriptions of cybersecurity tasks. Section 3 introduces the typical graph mining techniques used in cybersecurity tasks. Section 4 introduces the general process of graph mining for cybersecurity. Section 5 elaborates on cybersecurity solutions using graph mining techniques. Section 6 provides typical graph-based cybersecurity public datasets and toolkits. In Section 7, we point out future directions. Finally, Section 8 is a summary of this survey.

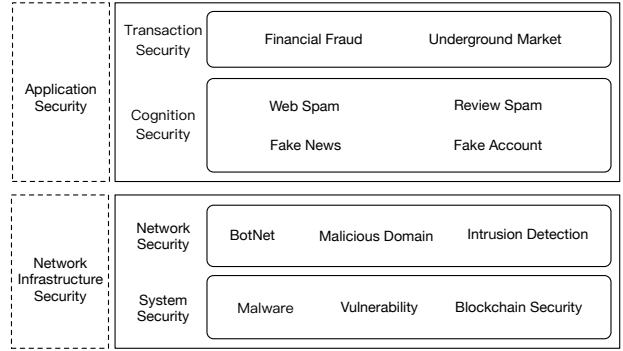

Fig. 1. Taxonomy of cybersecurity in this survey (the left two main categories and right four sub-categories). For each sub-category, we list typical tasks modeled by graph mining techniques.

## 2 BASICS OF CYBERSECURITY

Cybersecurity is the collection of tools, policies, actions, and technologies to protect cyber assets. These assets include users, network infrastructures, applications, systems, and the total transmitted and stored information in the cyber environment. Cybersecurity aims to ensure the availability, integrity, and confidentiality of these assets against cyber attacks [179]. Considering its importance, numerous advanced cyber defense strategies are proposed to ensure the safety of cyberspace. It's reported that the market for artificial intelligence in cybersecurity will grow from $1 billion in 2016 to a $34.8 billion by 2025 [169]. However, cyber-attacks are evolving and taking adversarial actions constantly, posing persistent threats to cybersecurity. For example, as a key threat carried out by attackers to breach cybersecurity, malware evolves rapidly to avoid detection due to the technologies of automatic generation of their variants, leading to an exponential growth of new malware samples [219]. Therefore, some significant tasks of cyber attack and defense should be modeled and continuously concerned by researchers. In this section, we first present the taxonomy of cybersecurity, then provide an overview of these cybersecurity tasks.

### 2.1 Taxonomy

The taxonomies of cybersecurity tasks are various from different perspectives. In this survey, we mainly focus on cybersecurity tasks that are extensively modeled by graph mining techniques. The detailed taxonomy of cybersecurity is depicted in Fig. 1. Specifically, we divide cybersecurity tasks into two main categories, namely, application security and network infrastructure security. Network infrastructure security focuses on protecting key infrastructures and components of the internet, such as Domain Name System (DNS), network traffic, and routers, to ensure that the internet works in a trustworthy environment [200]. We further classify network infrastructure security at the network and system level, which includes specific network and system security tasks (e.g., botnet detection and blockchain security). Application security aims to protect the security of various applications that run on top of network infrastructures, such as social media and financial businesses. It was further classified into cognition security and transaction security. Cognition security mainly focuses on the security issues affecting human cognition from various cyber applications (e.g., fake news and review spam), while transaction security emphasizes security issues in online transactions that threaten human assets and financial order (e.g., financial fraud and underground markets). It should be noted that most of the cybersecurity tasks can be abstracted as anomaly detection, which aims to detect rare occurrences in samples [4]. Since this survey focuses on specific cybersecurity tasks, we do not introduce anomaly detection separately but describe related methods in specific tasks.

## 2.2 Overview of Cybersecurity Tasks
Based on the taxonomy of cybersecurity, we give an overview of these cybersecurity tasks.

*2.2.1 Transaction Security.* Over the past years, the rapid developments of the Internet promoted the flourishing of online transactions and digital finance. However, it in return has become a ground for attackers to perform malicious transactions for profits, leading to huge losses of human assets and damaging the regular financial market operation. By means of the Internet, transactions can be performed both on the surface Internet and darknet, the latter of which is estimated to be much larger than the former. The transactions on the surface internet are regulated by governments and financial institutions. However, malicious transactions still continue to occur, which is known as financial fraud [128]. The transactions on the darknet form the underground market [41], which is unregulated and thus becomes an ideal platform to execute illegal transactions.

**Financial Fraud**. Financial fraud aims to obtain beneficial gains in unethical and illegal ways by exploiting rule vulnerabilities of the financial market. It is an essential issue that damages both individuals' daily life and the financial market and has become the main threat to transaction security. For instance, in the United States, fraud activities in the insurance field led to $300 billion in financial losses a year [128]. We mainly focus on four typical applications that are modeled by graph mining techniques, namely money laundering, catch-out, loan default, and insurance fraud.

**Underground Market**. The underground market can be defined as an illegal, untaxed, and unregulated transaction market running on a portion of the Internet referred to as the darknet [62]. Different from the surface Internet we used daily, the darknet is an encrypted Internet network that hides the IP address and uses virtually untraceable cryptocurrency (e.g., bitcoin), facilitating anonymous transactions [233]. Numerous illegal trades (e.g., drug trafficking, arms smuggling) happening in the underground market pose great threats to the financial order. For example, it is revealed that a group of cyber-criminals can make a profit of $864 million every year through renting Distributed Denial-of-Service (DDoS) attacks [234]. To combat these criminals, many crucial tasks are proposed based on underground forum (or hacker forum) analysis, such as key actor detection [235], account linking [233], illicit product identification [41], and Cyber Threat Intelligence (CTI) collection [143].

*2.2.2 Cognition Security.* Among the various content on cyberspace, there is mixed information that is false or with malicious intentions. This information propagates rapidly through social media, misleading people's conceptions and decision-making, and further damaging the democracy and economy of the country [57]. For example, a widely spread fake news that Barack Obama was injured in the explosion made market value evaporate 13 billion [244]. Cognition security refers to the potential impacts of information on human cognition, ranging from web spam, fake news, and review spam, to fake accounts, and has become a hot topic in recent years. The typical threats to cognition security are introduced as follows.

**Web Spam**. Web spam refers to hyperlinked pages on the Internet that are created with intention of misleading humans and search engines [58]. By content and link manipulation (e.g., inserting keywords into web content or deliberately connecting the website to a large number of other websites), web spam obtains a high rank returned by search engines, which deteriorates the product quality of search engine providers. Moreover, these high-ranked web spam often serves as means of spreading malware and fishing attacks [158], which is performed by the known *drive-by download attack*. That is, the users may pay attention to the web spam and be induced to click a redirection Uniform Resource Locator (URL) to access a compromised website. To accommodate these issues, the research on web spam detection has become an arms race, especially for search engine industries [148].

**Fake News**. Nowadays, the explosive growth of fake news threatens both political and public interests since the information involved in fake news influences general cognitive preferences [57]. Social media such as Twitter and Facebook has become an ideal platform to spread fake news. As a result, detecting fake news from social media has become a significant task for protecting cognition security. Note that there has been no common definition for fake news. In this survey, similar to the broad definition of [244], we define fake news as *Fake news is false news published by a social member*. Such a definition means the term *fake news* should be about the entire information ecosystem. Thus, we also incorporate some related conceptions (e.g., false information [88] and rumors [17]) as a part of fake news fields.

**Review Spam**. Online review is an important factor affecting people's attitudes towards items in current e-commerce platforms. Fraudsters utilize this resource to mislead people's buying decisions for reshaping their businesses and even execute financial fraud. There exist two kinds of review spam, i.e., individual and collective review spam. Collective review spam (also called review spammer group [196]) refers to a spammer group that colludes and collectively works together to control the reviews of items. Compared to individual review spam, collective review spam is more detrimental to e-commerce platforms and harder to be detected due to the more complex inter-group dynamics [32].

**Fake Account**. With the boom of social media, numerous users have registered online accounts for communication, sharing knowledge, and entertainment. However, there also exist masses of fake accounts (also called Sybils), which do not correspond to real users and are harmful to the trustworthiness of the social network. For example, they can weaken the credibility of the network if users begin to doubt the authenticity of profile information [205]. Attackers often create these fake accounts to execute fake social engagement for particular purposes, e.g, adding the number of page likes or fake Twitter followers to promote services and products [66], political astroturfing [134], and even manipulating voting results [19]. Therefore, detecting and removing fake accounts is imperative for protecting legitimate users and maintaining the credibility of social networks.

*2.2.3 Network Security.* The network consistently suffers security threats both from outside and inside, such as DDoS attacks, and malicious scanning. Network security focuses on protecting the entire network environment against network attacks and ensuring the reliable operation of network infrastructures. Various network security tasks provide a safe communication guarantee for the development of various activities on the Internet. Network traffic analysis is a fundamental approach to these tasks [137]. Therefore, we review three typical tasks based on network traffic analysis in network security, namely, botnet detection, malicious domain detection, and intrusion detection.

**BotNet**. A botnet indicates an infected network, with a large number of compromised devices, each of which is a bot and controlled by a botmaster. The activity of botnets is mainly divided into 3 stages. First, the criminals transmit malware and virus to compromised devices. Then, the criminals supervise a few botmasters to manipulate a mass of bots through Command and Control (C&C) channels. Finally, the bots launch distributed attacks on the network. A large portion of cyber issues, for example, DDoS attacks, trojans, and malicious bitcoin mining, are all related to botnets. Botnets have already become prevalent tools, which may be used to perform new cyber attacks in the future.

**Malicious Domain**. Domain Name System (DNS) is an import service on the internet, which maps domain names to their corresponding IP address. As a critical role in network communication, domains are also significant resources for criminals to launch cyber attacks, such as phishing websites, spam mail, and botnets. Detecting malicious domains has become an essential topic in network security. Even worse, attackers utilize advanced techniques (e.g., Fast-Flux and Domain-Flux) to improve their flexibility across the domain space. For example, the attackers cover up the real location of a malicious service by continually changing the mapping IP addresses [164]. These adversarial behaviors by attackers pose a great challenge to malicious domain detection.

**Intrusion Detection**. Intrusion Detection System (IDS) is a network security measure that proactively protects the specific network and system from illegal attacks. It collects and analyzes information from systems or networks and detects any behavior that attempts to destroy the integrity, confidentiality, and availability of computer resources, that is, to check whether there are violations of security policies and signs of attacks, and make corresponding action reactions. For an IDS, cyber threats come from both outside and inside. The outsider threats mainly refer to malicious behaviors in network communication and hosts, such as abnormal network traffic and vulnerability attack. Insider threats are mostly caused by malicious behaviors of internal employees of the organization, such as malicious logins [129], leading to a huge loss to the organization.

*2.2.4 System Security.* System security is an important part of network infrastructure security. It aims to protect various systems in cyberspace from inside and outside attacks or hidden dangers. As two key components of security systems, malware and vulnerability detection have become vital measures to ensure the safe, stable, and long-term operation of cyber systems. In recent years, as a newly emerged cyber system that realizes peer-to-peer value exchange and trusted data sharing, blockchain has also drawn much attention from both industry and academic communities, especially the accompanying security issues.

**Malware**. According to [219], malware (also known as malicious software or malicious code) refers to all harmful software programs, including viruses, worms, trojans, spyware, bots, rootkits, ransomware, and so on. These programs are utilized as weapons to launch cyber attacks, such as stealing information, compromising computers, and crippling critical infrastructures, leading to severe damage to the information systems. It's reported that a kind of malware may release the privacy data of victims or permanently block their access unless paying more than a $5 billion ransom in 2017 [40]. Typically, Portable Executable (PE) malware in Windows systems is the majority of malware samples and Android-based malware also occupies a large part [219]. Therefore, we mainly focus on these two kinds of malware. Besides, malware detection and malware family classification are two main tasks related to malware. We refer to these and other malware-related tasks together as malware analysis for convenience.

**System Vulnerability**. System vulnerabilities refer to defects or errors during designing software or operating system, which may be exploited by criminals to attack or control the entire system through the implantation of Trojans, viruses, etc. The number of system vulnerabilities has increased explosively in recent years. It's reported by MITRE Corporation[1] that there already have been 17,8581 records in Common Vulnerabilities and Exposures (CVE) until June 19, 2022. Exploring high-accuracy vulnerability detection methods has become an urgent problem for system security. System vulnerabilities usually manifest in program code, such as the well-known Structured Query Language (SQL) injection vulnerability. Therefore, the source code analysis is the basis of the vulnerability detection task [183, 208].

**Blockchain Security**. Blockchain is a decentralized and tamper-free system that establishes trust and achieves trusted transactions for the digital world. Based on the architecture of blockchain, many applications have proliferated in recent years, such as the well-known Bitcoin. Ethereum is one of the largest blockchain systems nowadays, whose emergence represents the arrival of the blockchain 2.0 era. Smart contracts are programs automatically running on top of Ethereum. It can be designed by developers to implement arbitrary rules for managing digital assets, making the automatic execution of contract terms possible [108, 109]. However, since the programming languages for the smart contract newly emerge, there may expose many vulnerabilities which are prone to be attacked. For example, In 2016, attackers exploited the reentrancy vulnerability of the Decentralized Autonomous Organization (DAO) contract and stole 3.6 million Ether [108]. Besides, due to the anonymity and high autonomy of blockchain systems, criminal activities occur consistently, such as money laundering and phishing scams, posing great risks to blockchain digital assets. Therefore, we mainly focus on graph mining solutions for two kinds of blockchain security threats: (1) vulnerabilities of the designed blockchain system itself (e.g., smart contract vulnerability). (2) Criminal activities in the blockchain (e.g, malicious users utilize the blockchain to execute money laundering).

---

[1]https://www.cve.org

## 3 OVERVIEW OF GRAPH MINING TECHNIQUES

In this section, we will first present some basic concepts of graph mining techniques. Afterwards, detailed taxonomy and descriptions of typical graph mining techniques used in cybersecurity are introduced.

### 3.1 Basic Concepts

Many real-world cybersecurity data can be naturally abstracted or manually constructed as a graph, aiming to model the relationships and complex interactions between cyber components. Formally, a graph is defined as follows.

DEFINITION 1. Graph. *A graph is denoted as $G = (V, E)$ where $V$ is the node set and $E$ is the edge set. A node type mapping function $f_\mathcal{A} : V \rightarrow \mathcal{A}$ and an edge type mapping function $f_\mathcal{R} : E \rightarrow \mathcal{R}$ are defined in the graph, where $\mathcal{A}$ is the node type set and $\mathcal{R}$ is the edge type set.*

The basic topological information of a graph can be depicted by some properties, which are also important features to identify cyber threats. For example, as a basic property to measure how tightly interconnected a graph is, *density* is also widely used in malware analysis and botnet detection. Here we list typical graph properties and their descriptions in Table 1. Note that these general properties can be used to characterize all kinds of graphs.

Table 1. Typical graph properties (structure-based statistical features) and their descriptions.

| Property level | Property | Description |
|---|---|---|
| Node-level | In/out (weighted) degree | The number (weight) of in/out edges associated with the vertex. |
| | Centrality | Measure node importance, including degree centrality, betweenness centrality, and closeness centrality. |
| | PageRank [150] | Measures the importance of a node. The basic idea is that the importance of a node is roughly determined by the number and features of links to it. |
| | Local clustering coefficient | Measure the degree of closeness of a node to its neighbors, which can be calculated by the actual number of edges between neighbors of a node divide the maximum number of possible edges between neighbors of the node [150]. |
| Edge-level | Shortest path | The shortest path length between node $i$ and node $j$. |
| Graph-level | Diameter [45] | The maximum length of the shortest path between any two nodes in a graph. |
| | Degree distribution [153] | The degree distribution $P(k)$ of a network is defined as the fraction of nodes in the network with degree $k$. If there are $n$ nodes in total and $n_k$ of them have degree $k$, then $P(k) = \frac{n_k}{n}$. |
| | Density | Measure how tightly interconnected a graph and can be defined as the average normalized degree. |

Broadly speaking, cybersecurity graphs in reality are mostly heterogeneous, such as traffic flow graph [81] and code property graph [208], where the nodes or edges in the graph contain multiple types. We define the heterogeneous graph and homogeneous graph as follows.

DEFINITION 2. Homogeneous/Heterogeneous graph [166]. *A graph is called* heterogeneous graph *if the node type set $|\mathcal{A}| > 1$ or the edge type set $|\mathcal{R}| > 1$, otherwise it is called* homogeneous graph.

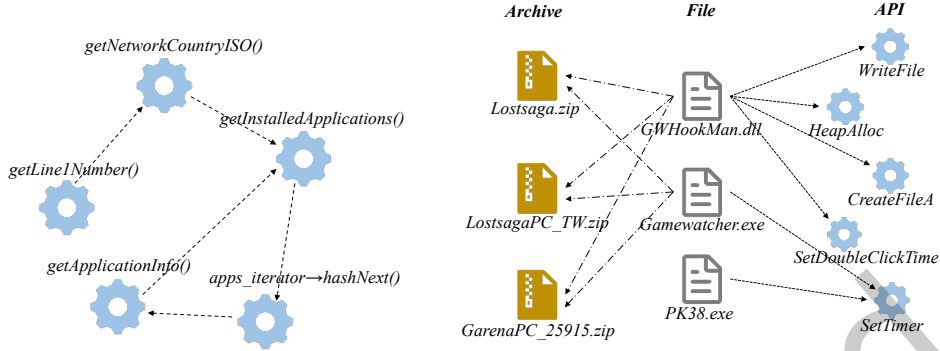

Fig. 2. An example of homogeneous graph (**left**) [125] and heterogeneous graph (**right**) [40]. The homogeneous graph is an API call graph of Android programs, where the nodes denote APIs and edges denote API call relations. The heterogeneous graph is a file dependency graph of Windows Portable Executable (PE) files. It includes multiple types of nodes (Archive, File, and API) and edges (e.g., Archives-include-Files and Files-call-APIs).

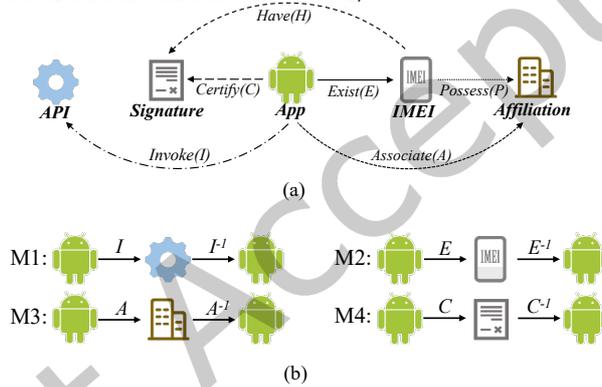

Fig. 3. Network schema (a) and meta-paths (b) for Android malware detection [218]. The symbol with superscript **-1** indicates inverse relation (e.g., $I^{-1}$ indicates "invoked-by").

EXAMPLE 1. Fig. 2 presents an example of the homogeneous/heterogeneous graph in the malware detection task. The homogeneous graph (**left**) is an Android API call graph where nodes denote APIs and edges denote call relations between APIs. The heterogeneous graph (**right**) depicts more complex interactions between system entities which include four types: Archive, File, API, and DLL (short for Dynamic Link Library).

For homogeneous graphs, the basic structural information can be captured by **first-order proximity** and **second-order proximity** [172]. The basic idea of first-order proximity is the nodes that are highly interconnected should be embedded closely together. For example, two Application Interfaces (APIs) with calling relationships may implement similar functions. However, in some cases, disconnected nodes may also have similar behaviors. Take the fraud detection task as an example, transactions between malicious accounts often use normal middle accounts as bridges. First-order proximity fails to model the similarity of these disconnected malicious accounts. To compensate for this limitation, second-order proximity characterizes the similarity of two nodes by comparing their neighbor structure, i.e., if two nodes have more common neighbors, their second-order proximity will be higher.

For heterogeneous graphs, the basic pattern can be represented as a **network schema** [167], which can be seen as a meta template reflecting node types and link relations. While the high-level semantic of the graph, which is proved more useful for downstream tasks, can be captured by meta-path [167].

DEFINITION 3. Meta-path [167]. *A meta-path $\mathcal{P}$ is a path defined on a network schema. The path can be denoted as $A_1 \xrightarrow{R_1} A_2 \xrightarrow{R_2} \cdots \xrightarrow{R_l} A_{l+1}$, which defines a composite relation $R = R_1 \circ R_2 \circ \cdots \circ R_l$ between node type $A_1$ and $A_{l+1}$, where $\circ$ denotes the composition operator on relations.*

We can simplify the representation of a meta-path $\mathcal{P}$ as $\mathcal{P} = (A_1 A_2 \cdots A_{l+1})$ if no multiple relation types exist between adjacent node types. For a relation $R$ between node type $A_i$ and $A_{i+1}$, which is denoted as $A_i \xrightarrow{R} A_{i+1}$, the inverse relation $R^{-1}$ holds naturally for $A_{i+1} \xrightarrow{R^{-1}} A_i$. Generally, $R$ is not equal to $R^{-1}$, unless $R$ is symmetric.

EXAMPLE 2. An example of meta-path designed for Android malware detection is shown in Fig. 3. Fig. 3 (a) depicts a network schema that includes five entity types and six relation types. Fig. 3 (b) presents four meta-paths extracted from the network schema. These meta-paths capture different semantic information of the graph. For example, the meta-path M1: $App \xrightarrow{I} API \xrightarrow{I^{-1}} App$ captures the semantics of two Android apps calling the same API, and the semantics in M4 denotes that the two apps are developed by the same developer.

## 3.2 Method Taxonomy

According to the graph mining techniques used in cybersecurity, we organize existing methods into two main categories. (1) Statistical features. The methods in this category are based on manually constructed statistical features, which are high-dimensional and sparse. Statistical features are further divided into structural and attributed features, depending on whether leveraging attributed information of the graph. (2) Graph embedding [52, 191]. Unlike statistical features, graph embedding transforms feature vectors from sparse to dense and automatically preserves structural and attributes information. We give taxonomies of graph mining methods from three perspectives: structural/ attributed (whether using attributed information), shallow/deep (whether using deep learning), and homogeneous/heterogeneous (whether focusing on the heterogeneous graph). These taxonomies can reflect the characteristics of different cybersecurity tasks from a comprehensive view. For example, some cybersecurity tasks (e.g., botnet detection) are sensitive to structural features while other tasks (e.g., fraud detection) pay more attention to attributed information; the underground market analysis often operates on a heterogeneous graph and the function call relations are often constructed as homogeneous graphs in malware analysis. The detailed taxonomy of graph mining techniques used in cybersecurity is given in Fig. 4.

*3.2.1 Statistical Features.* Most statistical features are manually constructed from the graph, reflecting the general statistical characteristics of graph structure and attributes. Some other statistical features are tailored for specific cybersecurity applications, which may be effective but hard to generalize. It is also proved that statistical features can achieve comparable or even superior performance in depicting some basic graph properties (e.g., counting substructures), compared with advanced models (e.g., Graph Neural Networks (GNNs)) [26, 48, 206]. We divide statistical features into two main categories, structural and attributed features, which are introduced as follows.

**Structural Features**. Structural features depict the basic properties of the graph and have been widely used in earlier cybersecurity solutions. For instance, in/out-degree is a salient feature to distinguish botnets; community detection algorithms play a crucial position in identifying community structural patterns of cyber threats [111, 187, 202, 208]. Here we list typical structural features and their descriptions in Table 1. Despite the automated learning by prevalent graph embedding technologies, structural features are still served as effective indicators to detect many potential threats [143]. Besides, some advanced cybersecurity solutions are also motivated by the statistical characteristics of the graph structure [102, 107].

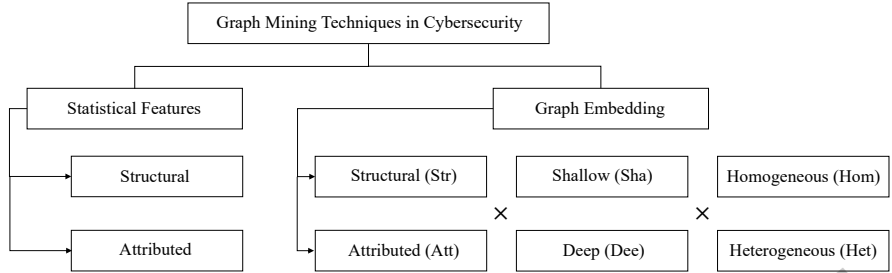

Fig. 4. Taxonomy of graph mining techniques in cybersecurity. The symbol ✗ indicates the Hadamard product which means that the sub-categories from different views can perform arbitrary combinations. Accordingly, a graph embedding method belongs to multiple sub-categories (e.g., Node2vec [54] belongs to Str+Sha+Hom embedding).

**Attributed Features**. Attributed features, such as node attributes and edge weights, are a significant complement of structural features, especially in the scenario of structural features being less discriminative. For example, in the smart home Internet of Things (IoT) vulnerability detection task, the traffic flow graphs may present similar structures, which is hard to identify vulnerabilities, while the attributes in the edges (flows) can reveal the vulnerability risks, such as "password" and "token" [78]. Some other attributed features are constructed for specific cybersecurity tasks. Take review spammer group detection as an example. The average review time and rating variance are two key attribute-based features for identifying spammer groups since spammer groups often post fake reviews in a short period and tend to rate similar scores [195]. Belief propagation (BP) algorithms are often utilized for passing this homophily between connected nodes [87, 170]. Besides, attributed features are often incorporated into graph embedding models and served as initialized node vectors.

*3.2.2 Graph Embedding.* Generally, assuming the data lie in a low dimensional manifold, graph embedding algorithms aim to not only reduce the high dimensionality of the non-relational data but also capture the structural and attributed information of the graph [15]. Each node is embedded into a dense vector and then performs downstream tasks. Thus from the view of utilized information, graph embeddings can be divided into structural and attributed. Besides, from the view of model architecture, many shallow methods (e.g., methods based on matrix factorization [209] and random walk [126]) and advanced deep learning-based methods (e.g., GNNs) are adopted to learn graph embeddings. Also, from the view of organization forms of graph data, the graph embedding methods focus on either homogeneous graphs or heterogeneous graphs. We divide graph embedding methods from these three views and the taxonomy on each view can cover all the graph embedding methods. Typical methods are introduced as follows.

**Structural vs. Attributed**. Structural embedding aims to preserve the structural information of graphs in embedding. Intuitively, if two nodes are connected or their local structures are similar, their node embeddings should have a closer distance. Thus, many works used random walk to sample local node sequences for node similarity learning. DeepWalk [126] incorporated the topological information into random walks and used Skip-Gram [115] to maximize the probabilities of neighboring nodes. LINE [172] further introduced weighted node distribution and proposed first-order proximity for modeling the joint probability of nodes with observed links. Node2vec [54] generalized DeepWalk by using a $2^{nd}$ order random walk procedure to sample node sequences and learned first-order proximity and second-order proximity simultaneously. To model the structural identity, Struct2vec [138] designs a weighted random walk traversing a hierarchical graph. Structural embedding methods are widely used in structure-sensitive cybersecurity tasks, such as function call graph structure analysis in malware detection [39, 125], pattern learning of fake news spreading within a social network [46], transaction behavior modeling in financial fraud detection [111], etc.

Unlike structural embedding only considers structural information, attributed embedding captures graph structure and node attribute information (e.g., fraudster profiles in social networks) simultaneously. [209] first proved DeepWalk is equivalent to matrix factorization and proposed text-associated DeepWalk (TADW) to incorporate node attributes into embedding. Many advanced graph embedding methods also support incorporating attributed information (e.g., GraphSAGE [60] and Graph Attention Network (GAT) [177]). Attributed embedding is more suitable for complex cybersecurity tasks where structural features are not distinguishable, such as threat analysis of dark net [233, 235] and system entity correlation modeling in malware detection [40].

**Shallow vs. Deep**. In the early stage, the majority of graph embedding methods are based on shallow models, whose trainable parameters are solely node embeddings [59]. These shallow models applied in cybersecurity are largely inspired by matrix factorization and random walk techniques. Matrix is a natural tool to denote graphs such as the adjacency matrix and Laplacian matrix. Therefore, we can easily obtain node embeddings by matrix factorization. Also, the learning node embedding can also reconstruct the whole graph structure by matrix multiplication [113]. Compared with matrix factorization which uses a deterministic way to learn node embedding, methods with random walk are more flexible and mostly included in structural embeddings, such as Node2vec, Deepwalk, and LINE.

The repaid progress in deep learning has prompted applying deep learning to graph embeddings, which are known as Graph Neural Networks (GNNs) [203, 241]. GNNs are deep models that update node embedding through neighbor aggregating and provide a more convenient and scalable way to learn graph embeddings. These deep models can be divided into spectral methods and spatial methods. Graph Convolution Network (GCN)[86] is a well-known spectral method, which generalized convolutions operation to graph data. Spatial methods, known as Message Passing Neural Network (MPNN) [50], define convolution operation based on the spatial relations between nodes. To capture the importance of messages from different neighbors, Graph Attention Network (GAT) [177] introduced attention mechanisms to the message-passing process. GraphSAGE [60] further adopted neighbor sampling to achieve scalability. It designs the mini-batch propagation to support inductive learning. To model temporal changes of the graph (e.g., network traffic graphs), Spatial-Temporal Graph Neural Network (STGNN) incorporates sequential models into the traditional GNNs, which is effective in some real-time applications (e.g., intrusion detection systems [27]).

**Homogeneous vs. Heterogeneous**. Homogeneous embeddings have been extensively studied for learning representations of homogeneous graphs. Many well-known embedding methods, from the random walk-based methods Node2vec, Deepwalk, and LINE, to the prevalent GNN methods, GCN, GAT, and GraphSAGE, are designed for homogeneous graphs. Quite a lot of these methods are based on homophily assumption, i.e., nodes with similar labels tend to connect with each other, and thus should have similar embeddings. Many cybersecurity tasks also hold this assumption [20, 145]. For example, the connected webs in the web link graph may have the same label (normal or spam). Correspondingly, these homogeneous graphs are often used to model correlations or similarities between nodes. Another type of homogeneous graph aims to model dependency relations, such as function call graphs for malware detection [71], network traffic graphs for abnormal traffic detection [33], etc. These graphs are not constructed based on homophily assumption, thus the goal is often to obtain embeddings for graph-level tasks.

To model the heterogeneity of graphs, many works focus on studying heterogeneous graph embeddings [191]. Semantic information is more significant in heterogeneous graph scenarios. For example, the two hosts who frequently send packets using the same protocol likely have similar purposes and thus should have similar representations [237]. Meta-path guided random walk was proposed to capture this semantic similarity [40, 227], which continues the walk along with pre-defined meta-paths. Besides, many advanced heterogeneous embedding methods rely on aggregating messages from heterogeneous neighbors, which mainly consist of two solutions. One is transforming attributes from different types of entities into the same space through type-specific mapping functions [107, 212, 225]. After obtaining target node representations through aggregating specific types of neighbors, multiple fusion methods (e.g., concatenation) can be applied to get the final node embeddings. Another solution is using meta-path guided neighbor aggregation [69, 188, 192]. A typical architecture of this solution is shown in Fig. 5. Note that the aggregation function has multiple choices. For example, it was designed based on the attention mechanism in Heterogeneous Graph Attention Network (HAN) [192], which is also a prevalent choice in many cybersecurity solutions [69, 180, 234]. Finally, the node representations under different meta-paths can be fused by multi-view fusion technologies [131, 192].

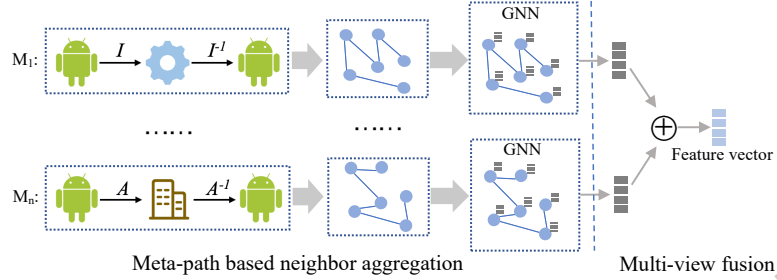

Fig. 5. The typical architecture of heterogeneous embeddings used in cybersecurity (taking malware detection as an example). It includes two procedures: (1) Meta-path based neighbor aggregation. Multiple homogeneous graphs (views) are constructed w.r.t. specific meta-path based neighbors, then perform neighbor aggregation in each view to obtain node representations. (2) Multi-view fusion. Different meta-path based node representations are fused to obtain the final feature vectors.

## 4 GENERAL PROCESS OF GRAPH MINING FOR CYBERSECURITY

The general process of graph-based cybersecurity solutions mainly consists of five steps. First of all, we should focus on one specific cybersecurity task (e.g., botnet detection) among various cyber attacks, and analyze task properties to give a clear task definition. Then, in terms of the complex cyberspace mixing with all kinds of entities, we should collect and process the data which support the task properties. Based on the task definition and processed data, the application-specific graph can be constructed, which captures rich interactions between related cyber entities. Next, an optimization model should be carefully designed for this task. Finally, the proposed model would be assessed by corresponding metrics and further deployed in real-world scenarios. In this section, we will introduce these crucial steps for designing effective cybersecurity solutions.

### 4.1 Task Definition

The step of task definition requires an in-depth understanding of the cybersecurity task, thus providing detailed task properties for designing proper models. Task definition mainly consists of two key steps: (1) Determining whether a cybersecurity task can be effectively modeled by graph mining techniques. Intuitively, if the real-world cyber entities or events present explicit graph structure (e.g., botnet [90, 237], transactions [27, 198]) or implicit relations (e.g., two domains queried by the same end host tend to highly correlated [94]), then the related task can be modeled by graph mining techniques. (2) Transforming a cybersecurity task into a high-level graph-related task (e.g., botnet detection can be abstracted as a node-level classification task [118]). Note that this process also provides a guide for graph construction. For example, in the node-level network traffic classification task, nodes denote traffic flow [239]; while if the task is modeled as graph-level, then nodes can denote packets, and the whole graph represents a traffic flow [153].

### 4.2 Data Collection

Cybersecurity data can be collected from various sources. In the domain of network infrastructure security, traffic flow and log records are two essential data sources since they reflect real-time network conditions or system behaviors. Besides, the passive DNS database (or other data sources, e.g., WHOIS records) stores the mapping relation between IP and domain. Various tools can be leveraged for traffic flow monitoring and collection, such as Wireshark and Netflow. The application security field includes more data sources, such as social media, hacker forums, source codes, etc. After we obtained these raw data from multiple data sources, many of them should be processed in advance for further utilization. For example, the raw traffic data is normally stored in pcap format, which should be further processed by network traffic analysis tools (e.g., tshark, Tstat); in malware detection, PE files need to be unpacked first by specialized tools (e.g., disassembler and memory dumper tools) if it is compressed. In addition to manually collected datasets, there are also many public graph-based cybersecurity datasets, we summarize them in Section 6.

### 4.3 Graph Construction

Based on task definition and collected datasets, the graphs can be constructed to explicitly model the internal relations of data. Generally, task definition gives a high-level guideline for constructing graphs and the collected datasets determine the node set and attribute set. The edges capture the relation between nodes, which should be carefully added based on priori knowledge and specific tasks.

In terms of the nature of a graph, we can divide the constructed graph into homogeneous and heterogeneous. Homogeneous graph construction is mostly based on explicit relations of data or manually designed metrics. For example, the graph whose nodes denote host and edges denote network traffic flow can models simple network communication behavior [187, 190]. The sensor graph in [142] is a KNN graph and is constructed based on the metric of pair-wise node similarities. Heterogeneous graph construction is more complicated. For example, the malicious domain detection task requires multiple entities (e.g., domain, IP, Account, CNAME record) and relations (e.g., map, query, register) [164]. In these cases, one can design the network schema first and then construct the concrete graph instance.

From the view of application scenarios, some graphs may only suit specific cybersecurity tasks. Many previous works have designed effective graph types for different applications. To avoid making it from scratch, we summarize typical graph types, the corresponding node and edge sets, and application scenarios in Table 2. We also give a coarse-grained graph-type taxonomy based on cyber entities (e.g., files and news). For example, the file dependency graph and file distribution graph are all classified into file relation graphs. Note that the Code Property Graph (CPG) proposed in [208] is a joint data structure merging AST, CFG and PDG, which therefore incorporates both of their node and edge types.

### 4.4 Model Design

The model design aims to design optimization models by deeply analyzing the task property and constructed graphs. The key idea is finding crime clues and choosing proper graph mining techniques to filter them. Traditional methods manually constructed discriminative graph features and fed them into ML classifiers. However, in some cases, suspicious patterns are hard to find, especially with the emergence of adversarial and more complicated crime behaviors. Therefore, graph embedding technologies have been proposed to automatically mine complex suspicious patterns. We can design these models depending on the task properties. For example, to capture the structural patterns of a botnet, random walk-based graph embedding methods can be used [118]; for the temporal patterns of Unmanned Aerial Vehicle (UAV) dynamic graphs, STGNN is considered more effective [27]. It is worth noting that traditional statistical analysis is still a fundamental means to motivate the design of advanced graph models [107].

Table 2. Typical application-specific graphs and their application scenarios.

| | Graph type | Node | Edge | Application scenarios |
|---|---|---|---|---|
| Web Relation Graph | Web Link Graph | Web | Web links | Web spam |
| | Web Redirect Graph | Web | Web redirection | Web spam |
| File Relation Graph | File Dependency Graph | System entity (e.g., File, API, process) | Dependency relation | Malware |
| | File Distribution Graph | Network entity (e.g., IP, File, domain,etc) | Network behaviors (e.g., request) | Malware |
| Network Traffic Graph | Host Graph | Host | Network traffic between host | Botnet detection; Intrusion detection |
| | Domain Resolution Graph | Domain/IP | Domain-IP mapping | Malicious domain |
| | Traffic Activity Graph (TAG) | Network entity (e.g., IP, Port, Protocol) | Multiple relations between network entities | Botnet; malicious domain; intrusion detection |
| | Flow Similarity Graph | Network traffic flow | Flow similarity | Network traffic classification |
| | Packet Sequence Graph | Packet with direction and length | Temporal dependency between packets | Network traffic classification |
| Code Graph | Control Flow Graph (CFG) | Basic blocks of a function | The execution order of basic blocks | Vulnerability |
| | Abstract Syntax Tree (AST) | Syntax structures of source code (e.g, identifier, expression) | Syntax relations | Vulnerability |
| | Program Dependence Graph (PDG) | Basic blocks of a function | Control and data dependence | Vulnerability |
| | Function Call Graph (FCG) | Function | Function call | Malware detection |
| | Code Property Graph | Incorporate AST, CFG and PDG | | Vulnerability; blockchain |
| News Graph | News Similarity Graph | News content (e.g., word, image) | Content similarity | Fake news |
| | News Propagation Graph | User behaviors (e.g., retweet, response) | Temporal order of user behaviors | Fake news |
| User Behavior Graph | Account-device Graph | User/Account/Device | User/Account logins device | Intrusion detection; financial fraud |
| | Social Media Interaction Graph | Social media entities (e.g., user, tweet, news) | User behaviors (e.g., retweet, response) | Fake news; fake account |
| | Review Graph | User/Review/Item | User review behaviors | Review spam |
| | Transaction Graph | Transaction entities (e.g., user, account, transaction) | Transaction behaviors | Financial fraud; blockchain |
| | Underground Forum Graph | Forum entities (e.g., user, thread, comment) | User behaviors (e.g., reply, comment) | Underground market |
| | User Similarity Graph | User | User similarity | Fake news |
| | Social Relation Graph | User/Account | Social Relation | Fake news; fake account |
| Others | User Interface Graph | UI components (e.g., ListView, Button) | Triggered events | Malware |
| | Power Grid Graph | Generators and Substations | Transmission lines | Vulnerability |
| | Vulnerability Dependency Graph | Vulnerability | Vulnerability dependency | Vulnerability |
| | Alert Graph | System alert | Alert correlation | Intrusion detection |
| | Sensor Graph | Sensor | Sensor similarity | Intrusion detection |
| | Unmanned Aerial Vehicle Graph | Unmanned Aerial Vehicle | Communications between vehicle | Intrusion detection |
| | Character Variation Graph | Character | Variation relation | Review spam |

## 4.5 Model Assessment

The effectiveness of a graph-based cybersecurity solution is measured by the model assessment procedure. Most of these models are classification-based. For example, malware detection is a binary classification task to classify a program as malicious or not, or a multi-class classification task to identify malware families. These classification-based models can be assessed by classification-based metrics (e.g., accuracy, Micro-F1). Particularly, IDS pays more attention to reducing false alerts, thus False Positive Rate (FPR) is an important metric to evaluate the effectiveness of IDS. Besides classification-based models, there are also some unsupervised models based on clustering, such as unsupervised malware family clustering [39], which can be assessed by clustering-based metrics (e.g., Adjusted Mutual Information (AMI)). Code similarity search is a fundamental procedure in malware detection and its assessment is similar to the recommender system, thus Rank-n and Mean Reciprocal Rank (MRR) can be used as evaluation metrics. In addition, the real-world deployment of graph-based models requires high time efficiency, which is often measured by running time [38, 207]. Visualization is also a significant model assessment means, especially when the task couldn't be evaluated by traditional metrics, such as analyzing different application components in mixed network traffic [81].

Table 3. Typical graph approach used in transaction security.

| Task | Graph Type | Graph Approach | Task Level | Paper |
|---|---|---|---|---|
| Financial fraud | Transactions | Attributed (statistical feature) | Subgraph | [76, 98] |
| | | | Node | [30] |
| | | Att+Dee+Hom (graph embedding) | | [111, 198, 199] |
| | | Att+Dee+Het (graph embedding) | | [69, 70, 240] |
| | Account-device | | | [102, 106, 107, 180] |
| Underground market | Underground forum | Str+Sha+Het (graph embedding) | Edge | [235] |
| | | Att+Dee+Het (graph embedding) | | [233] |
| | | | | [41] |
| | | | Node | [234] |
| | Word co-occurrence | Att+Sha+Hom (graph embedding) | Node/graph | [143] |
| | Transaction | Structural (statistical feature) | Node | [36] |
| | Host | Attributed (statistical feature) | Subgraph | [51] |

## 5 CYBERSECURITY SOLUTIONS USING GRAPH MINING

Graph mining techniques have been widely used in cybersecurity tasks. In this section, we summarize and elaborate on these graph-based cybersecurity solutions based on the taxonomy in Section 2.1.

### 5.1 Transaction Security

Among the strategies of transaction security protection, graph-based methods model behaviors and attributes of transactions simultaneously, which attract more attention recently. Typical papers and corresponding graph approaches are listed in Table 3.

*5.1.1 Financial Fraud.* Traditional methods most rely on prior knowledge or individual features but ignore the rich interactions between users. A general characteristic of financial fraudsters is community structure, i.e., the network containing fraudsters is often dense-connected, which motivates the application of graph mining techniques in financial systems (e.g., Alipay[2]) [106, 107, 111]. [111] extracted high-risk dense subgraphs and obtain the risk scores based on homogeneous graph embeddings for further screening. The aggregation and burst phenomenon observed by [107] suggested that a large number of fraudsters logging the same device in a short time tend to be high-risk. It initiated user node vectors by temporal activity statistics and utilized GNNs to aggregate local neighbors. However, normal accounts may also log in to the same device so that the above

---
[2]http://render.alipay.com/p/s/download

GNN-based models raise a high false positive rate. [106] solved this problem by introducing adaptive receptive paths. Besides these general solutions, graph mining techniques have also been successfully used in many specific tasks of financial risk management recently. We introduce these tasks and corresponding solutions as follows.

**Money Laundering**. Money laundering is a financial industry behavior, which makes illegal income legalization. It's reported that more than 150,000 people were killed by Mexican drug cartels since 2006; 700,000 people per year are sold in a human trafficking industry [198]. These illegal activities all rely on complex money laundering operations. The government and financial institutions have paid tremendous resources into anti-money laundering (AML) but few effects are obtained. Traditional Transaction Monitoring Systems (TMS) mainly rely on rule-based threshold protocols and scrutinizing by analysts, which suffer lower scalability and higher false positive rate. Graph model has emerged in AML recently since it captures cash-flow structures and account attributes in a uniform graph. One can use accounts [30, 198] or transactions [199] as nodes and edges denote cash-flows to model suspicious money laundering behaviors. [30] emphasized the importance of social network metrics (e.g., centrality scores) for identifying criminals. Scalability and efficiency are two key factors for designing graph-based AML models since the transactions and accounts are dynamically growing [98, 198]. Therefore, existing works utilize scalable and dynamic GNNs (e.g., fastGNN [198], EvolveGCN [199]) to perform AML. Besides, despite several manually constructed datasets [198, 199], public datasets for research in the AML community are still rare.

**Cash-Out**. As another major financial fraud, cash-out aims to gain profits through illegal procedures, e.g., buying pre-paid cards and re-selling them [69]. Conventional cash-out detection is mainly based on statistical features of an individual user, but seldom fully exploits rich interaction relation of uses [236]. Statistically, cash-out fraudsters tend to have consistent behaviors with the meta-path based neighbors. This observation motivates researchers to construct an Attributed Heterogeneous Information Network (AHIN) with accounts, merchants, and devices as nodes and propose multiple meta-paths to model the login and fund transfer relations [69]. The architecture of this heterogeneous embedding model is shown in Fig. 5. Note that similar architecture is also used in many financial fraud detection tasks [70, 180, 240]. Besides, considering the scarcity of available labeled data, [76] in turn detected dense subgraphs in a single-step transaction graph. It captured suspicious signals by a class of metrics from four perspectives: time, capital, cyclicity, and topotaxy.

**Loan Default**. The prevalent online credit payment services such as Ant Credit Pay of Ant Financial[3] nowadays facilitate people in transactions but breed new loan default risks. As a core component of online credit risk management, default detection has drawn much attention in recent years. It aims to predict whether a user could fall into default in the future. Traditional methods extract individual user-related features, e.g., user profiles, transaction history, and social relations. However, the related behaviors of defaulting are complicated and couldn't be fully modeled by only individual user features. Generally, there are below traits of defaulters that should be considered: (1) Intrinsic properties, i.e., the individual user-related features mentioned above. (2) Interactions with other entities. Defaulters often exhibit some suspicious interaction patterns with outsiders. For example, defaulters may have frequent fund transactions with others [70]. (3) Adversarial behaviors. some defaulters may deliberately construct complex behaviors such as making a long fund transfer path [240]. To capture these characteristics, one can also use AHIN to model these entities as well as the complex interactions between them. Based on AHIN, typical heterogeneous graph embeddings are applied [70, 180, 240]. [240] designed multiple meta-paths to capture semantic correlations between entities, and multi-view information fusion technology was applied for aggregating different types of neighbors [70, 180].

**Insurance Fraud**. Insurance fraud indicates that the insured deliberately conceals the actual situation, induces the insurer to accept insurance, and then swindles the insurance money. Insurance fraud not only reduces the profits of insurance companies but also affects the social and economic benefits. It's reported that the Canadian automobile insurance fraud led to an estimated loss of 542 million Canadian Dollars in 2007 [194]. Recently, graph mining technologies have been proven to be useful tools to combat this fraud. Taking Alibaba's return-freight insurance fraud as an example, [102] found device-sharing graph can effectively discriminate between normal and fraud accounts and utilized GeniePath model [106] achieving the best performance compared with graph embedding and machine learning methods. The application in the real world showed that more than ten thousand dollars can be saved every month.

---

[3]https://www.antgroup.com/business-development?tab=finance

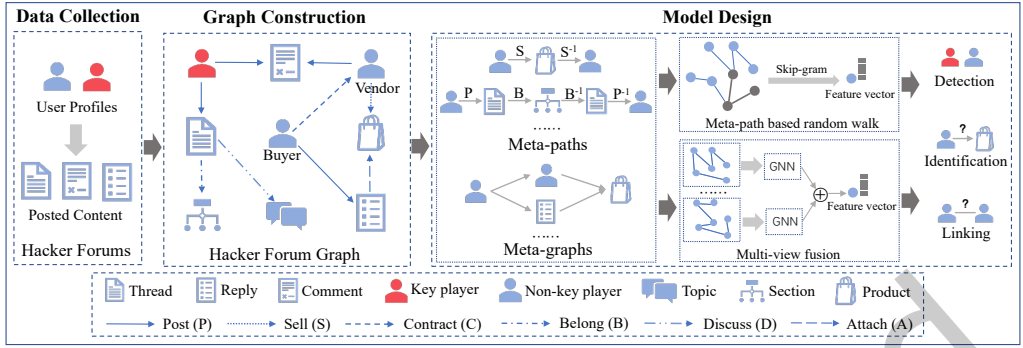

Fig. 6. The typical workflow of using hacker forums to perform underground market analysis. First, an AHIN (hacker forum graph) is constructed based on user profiles and their posted content in hacker forums. Then multiple meta-paths or meta-graphs are proposed to model semantic correlations between hacker forum entities. Next, graph embedding methods (e.g., meta-path based random walk, neighbor aggregation, and multi-view fusion) are utilized to obtain feature vectors. Finally, the feature vectors are used to perform three crucial tasks, namely, key player detection, production identification, and account linking.

Besides the above typical tasks using graph mining in financial fraud, there are many others that we haven't listed (e.g., online auction [123] and retail holding [176]). Nowadays, graph-based abnormal detection has become a typical approach to financial fraud detection. However, most approaches heavily rely on domain knowledge [128], leading a poor performance on adaptation and generalization. It's suggested that researchers endeavor to design graph-based solutions without any domain or expert knowledge, e.g., identifying abnormal nodes/subgraphs automatically in an unsupervised manner.

*5.1.2 Underground Market.* To combat criminals, security analysts need to continuously observe the latest status and changes of underground market activities [235], which is time-consuming and labored. Naturally, the various entities (e.g, vendor and buyer) and their interactions (e.g., developing or selling products) in hacker forums can be modeled as a heterogeneous graph. As a result, recently, researchers begin to utilize AHIN to model user activities (e.g., comments and posts) in the hacker forums and propose various attributed and heterogeneous embedding methods to solve crucial tasks in underground market analysis [41, 233, 234]. The typical workflow of these methods is depicted in Fig. 6.

As can be seen from Fig. 6, there are three crucial tasks in underground market analysis. The first task is key actor detection, which was first proposed in [235]. Key actors indicate a group of users who play a vital role in the value chain (e.g., have the ability to initiate cyber criminals). Therefore, it is important to identify key actors in hacker forums. Toward this goal, only analyzing posts published by users is not enough, the interactions between users are more important. IDetective [234] is an automatic system to identify key players in hacker forums. It constructed an AHIN to model complex interactions among various entities in hacker forums. Based on AHIN, meta-path guided neighbor aggregation and multi-view fusion are applied to obtain the final user representation. Another essential task for regulating trade activities is linking different accounts to the same individuals [233]. Criminals often possess multiple accounts in the underground market, and linking them automatically is necessary for tracking criminals. The account linking task of drug traffickers was first addressed by [233]. It modeled account linking as a sim-supervised link prediction task (i.e., whether there exists a link between two accounts) and used a meta-path based AHIN model to obtain the similarity of two accounts. Identifying illicitly traded products in private contracts in underground markets is also an essential and challenging task [41]. Although the majority of transactions are private in hacker forums, the information of the vendor and buyer in a transaction is accessible. Based on this observation, one can leverage user profiles and their posts in hacker forums to identify the traded products [41].

Besides modeling user activities in the hacker forums as a heterogeneous graph, some other graph types are constructed to perform underground market analysis. [143] constructed a dynamic word co-occurrence graph based on hacker forums to proactively analyze emerging new threads. Since the hacker forum is also an important source of Cyber Threat Intelligence (CTI), through monitoring the evolution of word embeddings, experts can pinpoint emerging threats in terms of popularity and functionality. [36] explicitly modeled the drug trade between buyers and vendors in a transaction graph. Structural statistical features (e.g., density and centrality) and community detection analysis are utilized to mine the structure of criminal affiliations and collaborations. [51] constructed a host KNN-graph whose edge weight is the cosine similarity between host embeddings. A graph-based clustering method is proposed to find suspicious collaborative subgraphs.

Currently, hacker forums are primary data sources for graph-based underground market analysis. Another important source is network traffic. However, due to darknets hosting neither production services nor client hosts [51], the information for analyzing the traffic is limited. Several studies have presented their efforts to analyze the traffic of darknets using graph mining techniques [51]. Besides, graph mining techniques are mainly utilized to model financial activities in the underground market while some other worse activities (such as murder and terrorism) still haven't been well-addressed. Graph mining techniques are promising in mining inconspicuous clues of these untraceable activities.

*5.1.3 Summary.* Extensive studies have proved that graph-based methods especially AHIN have great potential for solving transaction security tasks. Considering the nature of transactions, three aspects should be concerned when designing a graph-based solution: (1) Rich attributes. The rich attributes of entities and relations in transactions are important to identify criminals [240]. How select useful attributes and incorporating them into the graph model is a challenging task. (2) Scalability. The graph size related to transactions is large and increasing over time. Therefore, the focusing points are not only the model effectiveness but also the memory-efficient graph representations [198]. (3) Robust. For emerging adversarial behaviors by fraudsters, robust graph-based solutions for transaction security are urgently required.

## 5.2 Cognition Security

Traditional methods for protecting cognition security mostly rely on content analysis (e.g., linguistic feature extraction) and characteristics of participants (e.g., identity and review time) [65, 244]. However, they ignored the rich interactions between participants and content. In recent years, graph-based methods have been proposed to capture these interactions and proved more effective and robust in identifying malicious information [95, 120]. In this section, we introduce graph-based solutions for cognition security tasks, which mainly include four categories, namely, web spam, review spam, fake news, and fake accounts. Besides, we also summarize the main characteristics of these methods in Table 4.

Table 4. Typical graph approach used in cognition security.

| Task | Graph Type | Graph Approach | Task Level | Paper |
|---|---|---|---|---|
| Web spam | Web link | Structural (statistical feature) | Graph | [171] |
| | | | Node | [20, 160] |
| | | Attributed (statistical feature) | Node | [1, 144] |
| | Web redirect | | | [154, 162] |
| Review spam | Review | Structural (statistical feature) | Subgraph | [215] |
| | | Attributed (statistical feature) | | [195, 196] |
| | | | Node | [135, 146, 181] |
| | | | Edge | [3] |
| | | Str+Sha+Hom (graph embedding) | Subgraph | [32, 83, 133] |
| | | Att+Dee+Het (graph embedding) | | [151] |
| | | | Node | [95] |
| | Character variation | Att+Sha+Het (graph embedding) | Graph | [80] |
| Fake news | User similarity | Str+Sha+Hom (graph embedding) | Node | [202] |
| | | Att+Dee+Hom (graph embedding) | | [110] |
| | News similarity | Attributed (statistical feature) | | [56] |
| | | Att+Dee+Het (graph embedding) | Graph | [193] |
| | News propagation | Structural (statistical feature) | | [189] |
| | | Att+Dee+Hom (graph embedding) | | [10] |
| | Social media interaction | Structural (statistical feature) | Node | [211, 243] |
| | | Attributed (statistical feature) | | [155, 156] |
| | | | Graph | [82] |
| | | Att+Sha+Het (graph embedding) | Node | [46] |
| | | Att+Dee+Het (graph embedding) | | [120, 223] |
| Fake account | Social relation | Attributed (statistical feature) | | [12, 14, 18, 224] |
| | | Structural (statistical feature) | | [145] |
| | Social media interaction | | Subgraph | [66] |
| | | Attributed (statistical feature) | | [9, 100] |
| | | Att+Dee+Het (graph embedding) | | [214] |
| | | | Node | [97] |

*5.2.1 Web Spam.* Existing web spam detection methods can be divided into two categories: content-based and link-based. Content-based features include hosts and domains of websites, source codes of webpages, etc. However, sometimes there are only a few contents on the websites so malicious pages are almost the same as benign pages except for some malicious URLs. Link-based methods identify web spam by analyzing web link structure, thus becoming an indispensable compensation of content-based methods. Concretely, *web link graph* [155] and *web redirect graph* [235] are the two mostly used graphs to model structural characteristics of web spam.

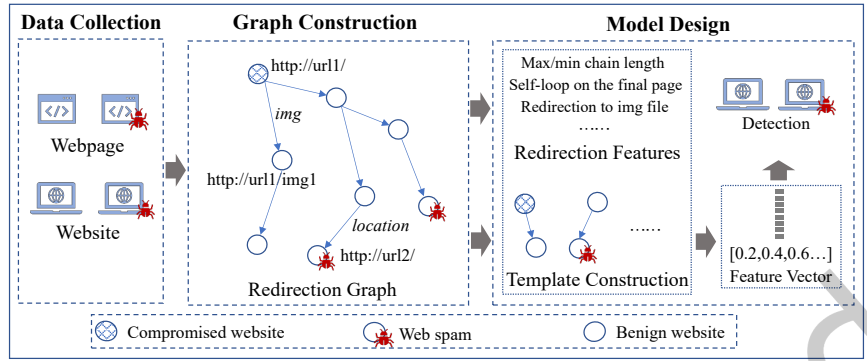

Fig. 7. The typical workflow of using redirection chain to identify web spam. The malicious redirection chain starts from a compromised web and ends with web spam. Based on the redirection graph, redirection features and benign/malicious redirection templates are constructed to form the feature vector for web spam detection.

**Low-quality Webpage**. For webpages with high rank and low quality, an observation is that the PageRank-based algorithms for search engine ranking highlight the degree of a website, making criminals add masses of hyperlinks referring to the target websites to improve rankings. Based on this observation, researchers constructed the web link graph to analyze the structural patterns of web spam. [8] extracted some structural statistical features such as in/out-degree and PageRank value from web link graph and performed node-level web classification. Content-based features are further incorporated by [144]. [20] found that linked websites tend to belong to the same class (malicious or benign), thus applying the clustering algorithm to refine the classification results. [1] enriched web link graph by introducing attributed statistical features (e.g., weighted links) and characterizes the similarities between websites by adding a graph regulation term. Considering the different linking conditions, [185] further adjusted the weights of the links adaptively according to the link conditions. The results showed that this method effectively reduced the ranks of spam pages.

**Drive-by Download Attack**. Web link graph was also used to detect web spam that performs drive-by download attacks. [160] analyzed the topology and frequency of web spam, and found web spam with different types of neighbor structures (e.g., multiple landing sites link to a single distribution site). [171] extracted structural features from the web link graph and used them to perform classification. To perform drive-by download attacks, redirection is a common technique used by criminals because multiple redirections can make detection harder. Compared with content-based features, redirection chains are more robust since they are not easy to obfuscate [162]. However, distinguishing redirections between benign webs and spam is a challenging thing. [162] firstly found that the web redirection graph of spam is different from benign webs in attribute and structure. To capture this discrepancy, some discriminative statistical features (e.g., country diversity and self-loop on the final page [162]) or redirection chain templates [154] are constructed. A typical workflow of these methods is depicted in Fig. 7. It extracted redirection chains from the web redirection graph and constructed benign and malicious redirection chain templates. The similarity vector between unknown redirection chains and existing templates was fed into a classifier to detect web spam.

Despite the great successes achieved in web spam detection, search engine industries are still making great efforts to deal with this threat nowadays. Link-based methods are widely applied and proven to be indispensable for web spam detection, especially when web contents are unavailable in some scenarios. However, these methods only capture some relatively simple structures which are easy to evade by smarter crimes. More complex structures and semantic features of malicious webs should be explored in the future, e.g., the semantic relations between webs in a meta-path.

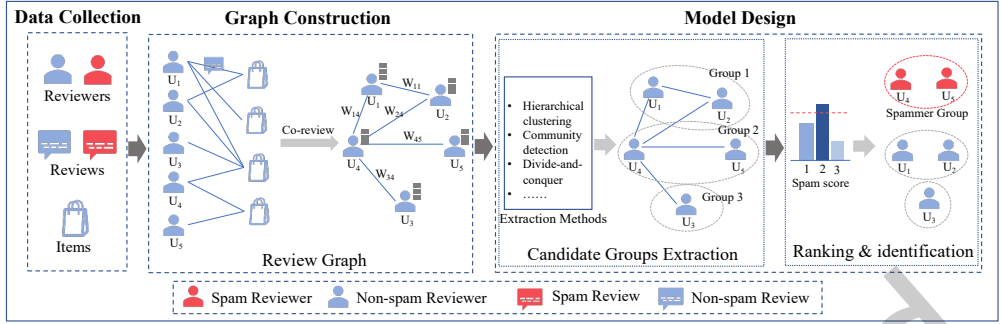

Fig. 8. The typical workflow of collective review spam detection using graph mining techniques. A co-review graph is deduced from the review graph, with attributed reviewers as nodes and collective degrees as edge weights. Based on the co-review graph, various graph-based methods (e.g., community detection) are used to extract candidate spam groups with high collective degrees. Each candidate group will be assigned a spam score to perform group ranking and spam group identification.

*5.2.2 Review Spam.* The attempts for detecting spam reviews/reviewers can be roughly divided into two categories: content-based and behavior-based. Content-based methods directly find clues in the reviews that fraudsters left. Early spam reviews show duplication characteristics, thus many studies focus on content similarity mining methods. To combat spammers that use more complex methods to generate reviews, many methods constructed manually-craft features of contents (e.g., the reviewer's sentiment) as input, and then perform text classification using machine learning algorithms. However, these content-based methods suffer performance decay under the adversarial actions of reviews. For example, spam reviews camouflage themselves as benign ones by replacing some characters [79]. Behavior-based methods model the interaction between reviewers, reviews, and items, requiring no inspection of content, thus are more robust. Current behavior-based methods mainly focus on detecting two kinds of review spam, i.e., individual review spam and collective review spam.

**Individual Review Spam**. Naturally, online review interactions can be abstracted as a heterogeneous review graph with reviews, reviewers, and items as nodes [181], which is a basis for individual review spam detection. Various attributed statistical features were proposed based on review graphs. [181, 182] first gave calculation methods for the trustiness of reviewers, the honesty of reviews, and the reliability of items. [3, 146] further used belief propagation and clustering methods to obtain abnormal scores of these entities. [135] considered both content and behavior features and proposed a unified framework for spam review detection. They also incorporated meta-data of reviews, such as the reviewer's identity, the time of posting reviews, and the rating of reviews. Some new content and structural features were proposed in [121], with a multi-iterative spam detection algorithm to improve the accuracy and reduce the complexity. To explicitly model the relations between reviews\reviewers and capture deeper semantic information of spam activity, [95] proposed a GNN-based Anti-Spam (GAS) model including a heterogeneous review graph and a homogeneous graph of comments, with high scalability and robustness.

**Collective Review Spam**. The key factors for describing the review spammer group are review time interval and rating score deviation [195]. Therefore, most existing works constructed the co-review graph whose nodes denote reviewer and edges denote collective behaviors (e.g., having similar ratings for items within a short time window). Based on the co-review graph, the typical workflow for review spam group detection is first cutting the graph into subgraphs (candidate spam groups), then ranking and identifying spam groups, which is depicted in Fig. 8. Specifically, these methods can be divided into (1) structural statistical features. [215] first extracted suspicious subgraphs which do not conform to common structural patterns of social networks, then performed hierarchical clustering to detect spam groups and their nested hierarchy. (2) attributed statistical features. The co-review graphs can be further enriched by node or edge attributes. For example, the edge weight was set as the number of common products reviewed [196]; the review time interval and rating score deviation were considered when designing spamming indicators [195, 228]. (3) Str+Sha+Hom graph embedding. It mainly uses random walk-based methods (e.g., node2vec) to learn latent structural patterns of reviewers [32, 83, 133]. The learned embeddings are further used to perform clustering to mine spammer groups. (4) Att+Dee+Het graph embedding. The feature vectors of reviewer nodes are initialized by reviews of this reviewer [151]. It proposed a GraphRNN framework to capture the long-range dependency of reviewers.

Both incorporating content and behavior features of reviews are the most effective and robust in current review spam detection systems. While graph mining technologies have been successfully applied to behavior modeling, few works model the content of reviews as a graph. For example, some variations of characters can be modeled as a character variation graph [80], which can deal with the camouflage of spammers. Also, the content can be presented in different languages. The graph structure can link the same meaning of different languages, which may be a potent tool for multilingual review spam detection.

*5.2.3 Fake News.* Traditional fake news detection methods include manual fact-checking and evidence-based approaches, which need external fact-checkers or knowledge databases for help. Then machine learning algorithms are used to detect fake news automatically based on content-based features but ignore the propagation patterns of fake news. Early studies for fake news propagation are based on social network analysis [28, 175]. They focus on quantitative analysis of the nature and characteristics of news dissemination in the social network. The state-of-the-art models currently are mostly based on user-news interaction graphs since they capture interaction features between users and news, which serves as a significant part of the nowadays fake news dissemination process. According to the above different focusing aspects, we divide existing graph-based fake news detection methods into three main categories, which focus on the analysis of user relations, news relations, and user-news interactions respectively.

**User Relation Analysis**. The underline social relations between users are useful for fake news detection. Intuitively, fake news may be quickly spread in a community structure. Therefore, [202] learned structural embeddings of user nodes based on community detection algorithms in a given social graph. [110] further considered the privacy of users in the real world. It argued that social relation graphs such as the following relationship of tweeters belong to user privacy and can't be utilized by models. Thus it constructed a fully-connected user graph where edge weights denote similarities of user traits. The node embeddings are trained using GNNs to capture potential relationships between users. These methods only use the graph to capture user relations, but use sequential models to capture behaviors between users and news.

**News Relation Analysis**. News relation analysis includes two aspects, intra-relation analysis, and inter-relation analysis. The intra-relation analysis captures the content dependency relation within a piece of news, while the inter-relation analysis aims to capture the similarities between different news contents. For intra-relation, [193] incorporated multi-modal information of news into a unified graph, including words, objects in news pictures, and entity conceptions from external knowledge bases. They used Point-wise Mutual Information (PMI) as edge weights to model the long-distance dependencies between words. Then a heterogeneous embedding model was applied to obtain the graph representation to perform fake news detection. For inter-relation, an attributed statistical method was proposed by [56]. It constructed a news KNN-graph based on the L2 metric, then performed belief propagation to obtain the labels of unknown news. Similar to user relation analysis, news relation analysis also failed to consider the rich interactions between users and news.

**User-news Interaction Analysis**. Instead of modeling user relations and news relations independently, current advanced methods consider interactions between users and news. The intuition is that the user behaviors toward news, such as comments and forwards, can reflect the trustiness of users and news. [82] collected tweets about a piece of news and construct a tweet graph. A credibility propagation method is applied to obtain the credibility of the news. Other methods modeled behaviors between user and news through capturing the news propagation patterns [10, 189, 243]. They pre-defined or automatically learned fake news propagation patterns [10, 189]. Recently, to explicitly model the interactions between users and news, user-news interaction graphs were constructed. [156] explored the interactions of publisher-news, user-news, and user-user. It proposed a tri-relationship fake news detection framework. Some works further captured more subtle hierarchical relations among interactions [155, 223]. [223] used attention-based heterogeneous embedding to capture local correlations between source news and forwards, as well as global correlations between news and related users. The micro-level comments graph and macro-level news propagation graph were constructed separately in [155]. [120] found the engagement temporality of users is distinctive for fake news, thus adding stance nodes to the user-news interaction graph, and using a sequence model to capture this characteristic.

Despite the extensive graph-based solutions for fake news detection, there also exist the following concerns. (1) Interpretability. Interpretability is highly needed in fake news detection since it can enhance the trust of the public [110]. More causal factors for identifying fake news are expected mined by graph mining techniques. (2) Unsupervised. The facticity of news is highly time-dependent, thus fake news detection datasets mainly depend on manual annotation which is time-consuming and costly. Unsupervised fake news detection methods are emerged in recent years but suffer relatively low accuracy [46, 211].

*5.2.4 Fake Account.* There exist two research directions for fake account detection, account attribute analysis, and social network analysis. The former focuses on collecting the attributed characteristics of fake accounts, such as account profiles and released information. However, the methods are infeasible when attributed information is unavailable [100]. Social network analysis models various social entities as a graph, with nodes denoting entities (e.g., accounts and posts) and edges denoting user behaviors (e.g., friendships and retweets), which focuses on capturing the abnormal structural patterns of fake accounts. We introduce these social network analysis methods as follows.

In the early days, fake accounts own some simple structural traits. Many attributed statistical features are proposed to incorporate these structural traits. [18] found that the connections between fake and normal accounts are often sparse in social networks. Therefore, a random walk sequence starting from normal accounts most likely land at normal accounts than fake accounts, and vice versa. They proposed a trust propagation method to model such probabilities and rank them to obtain suspicious accounts. The lockstep behavior is also a typical trait of fake accounts, where a group of fake accounts links to another account at around the same time (e.g., a sudden increase in page likes) [9]. An iterative clustering method is proposed by [9] for identifying groups of fake accounts. [100] found the fake accounts present a modest degree distribution both in scope and scale. They utilized local spectral graph diffusion to perform clustering near the seed nodes, achieving scalable searching in large social networks. [93] analyzed the structural and content characteristics (e.g., network density, user profiles, etc.) of social networks on Twitter. Three fine-grained classes of fake accounts are identified based on these characteristics.

To evade detection, smart fraudsters may present camouflage behaviors. For example, they add links to popular items or famous stars so that look similar to normal accounts [66]. To tackle this adversarial phenomenon, some works focus on the early detection of fake accounts, i.e., detecting these accounts when they just appear (e.g., at the time of registrations) and adversarial actions haven't been executed [14, 103, 224]. [14] found some distribution characteristics when new fake accounts sent requests to others. The probability of a new account being fake is obtained by (1) the choices of requesting others and (2) the targets' responses. [103, 224] focused on analyzing the characteristics of fake accounts when registering. They extracted synchronized features (e.g., fake accounts often have the same IP) and abnormal features (e.g., geo-location inconsistency). Based on these features, an account similarity graph is constructed and the dense subgraphs are identified as fake accounts. Other works focus on robust fake account detection [12, 66, 145, 214]. Early methods aiming to find dense subgraphs may ignore adversarial small-scale groups. To accommodate this, [66] proposed some robust metrics (e.g., edge density) and [145, 147] instead detect nodes with poorly reconstructed degrees. [12] relaxed the assumption of sparse connection proposed in [123]. It used account-level features to obtain edge weights and then performed random walks preferring to higher-weight paths. Attributed heterogeneous graph embeddings were used in [97, 214], which incorporated more node types (e.g., comment and hashtag) and edge types (e.g., post and reply), making it hard to carry out camouflages. [214] further proposed a reinforcement learning-based neighbor search mechanism to perceive the camouflages of new fake accounts.

Fake account detection based on social network analysis has been well-studied nowadays. However, only a few works focused on detecting fake accounts just appeared [14, 103, 224]. Since the detection is harder and the damage is worse over time, early detection or prevention is necessary, which is also the concerning direction of the present work. However, the information on new accounts is limited, thus it is possible to utilize cross-media information based on graph mining techniques. Besides, current graph mining techniques mainly focus on detecting fake accounts on popular social media, such as Twitter and Facebook. With the development of smart cities, fake accounts have eroded intelligent transportation, IoT, and other newly emerged areas, which may cause dramatic damages [61]. Graph mining techniques are promising to detect fake accounts in these more complex environments.

*5.2.5 Summary.* Graph-based solutions for cognition security provide a unified framework considering comprehensive features, including linguistic information and propagation patterns, user characteristics and behaviors, and external knowledge such as the background of news and social network. With more intelligent means of cognition attack [213] (e.g., camouflage and social engineering attack), cognition security should be enhanced by means of interdisciplinary knowledge, such as cognition science, psychology, and neuroscience [57]. In this regard, graph-based solutions for cognition security are still in the early stage.

## 5.3 Network Security

Network security has been widely addressed by network traffic analysis [114, 230]. Generally, a traffic packet can be represented by a 5-tuple {source IP, target IP, source port, target port, protocol}, and a traffic flow consists of multiple packets with the same 5-tuple. The well-known Deep Packet Inspection (DPI) technology analyzes traffic by inspecting the payload of packets. However, payload-based methods can't be used in encrypted traffic. To overcome this limitation, header features (e.g., IP, packet length, protocol) and statistic features (e.g, average duration, maximum inter-arrival time) are utilized to extract malicious features. However, both of the above methods ignore the graph structural features of traffic. Naturally, the network traffic components (e.g., traffic flow, domain, IP, etc.) form an enterprise network, or the whole Internet can be abstracted as a *Network Traffic Graph (NTG)*. Compared with traditional methods, graph mining methods based on NTG can avoid inspecting payloads and facilitate network security analysis [33]. In this section, we explicitly illustrate graph-based solutions for three main network security tasks, namely botnet detection, malicious domain detection, and intrusion detection. The characteristics of these solutions are summarized in Table 5.

*5.3.1 Botnet.* In the early days, botnets are centralized and the main goal is to detect centralized C&C servers. Therefore, many works analyzed C&C server activities based on traffic flows [55, 141, 161]. Later decentralized P2P-based botnet emerged to circumvent detection. Furthermore, botnets started to use more flexible C&C channels or mimic the communication patterns of normal hosts [187, 190]. As a result, traditional flow-based methods are not enough to characterize the botnet. Naturally, the centralized botnet presents a relatively simple graph structure, while decentralized is more complicated. This observation motivated researchers to study traffic topologies of botnets [118, 187, 190, 237].

Table 5. Typical graph approach used in network security.

| Task | Graph Type | Graph Approach | Task Level | Paper |
|---|---|---|---|---|
| Botnet | Host | Structural (statistical feature) | Node | [118, 186, 187, 190] |
| | | Attributed (statistical feature) | | [150] |
| | | | Graph | [90] |
| Malicious domain | Traffic activity | Att+Dee+Het (graph embedding) | Node | [237] |
| | | Structural (statistical feature) | | [75] |
| | | Str+Sha+Hom (graph embedding) | | [94] |
| | | Att+Dee+Het (graph embedding) | | [164, 165] |
| | Domain resolution | Structural (statistical feature) | | [84, 117] |
| | | Att+Sha+Hom (graph embedding) | | [64] |
| Intrusion detection | Host | Structural (statistical feature) | Subgraph | [33] |
| | Account-device | | Node | [45, 129] |
| | Alert | | | [122] |
| | Traffic activity | Attributed (statistical feature) | Subgraph | [78] |
| | Attack | | Graph | [231] |
| | Sensor | | | [142] |
| | Unmanned Aerial Vehicle | Att+Dee+Hom (graph embedding) | Node | [27] |

**Botnet Topology Analysis**. Many structure-based statistical approaches were proposed based on analyzing the topological traits of botnets. Due to bots need to communicate with many nodes, one obvious topological trait is that botnets usually present a dense intra-connection structure [29, 150, 186, 187]. In/out-degree is the most straightforward statistical feature to depict this characteristic [29, 150, 186]. [190] held that the normal nodes' degree follows a power law distribution and used the local ego-net distribution to identify bots. [187] used a modularity-based community detection algorithm to find dense connection structures, and cut the whole graph into botnet nodes and normal nodes. [186] went a step further. It first identified pivotal nodes (a set of highly interactive nodes), and then proposed a refined modularity-based community detection algorithm. The intuition is that bots may have a dense connection with pivotal nodes and a sparse connection with normal nodes. Another topological trait of botnets is their fast-mixing nature [118], i.e., the graphs of botnets can reach stationary distribution in a relatively short time. Therefore, [118] performed a uniform random walk in the host graph, then used the deviation between obtained node probabilities and stationary distributions to distinguish bots.

In addition to observed structural information such as in/out degree, some implicit structural differences between botnets and normal structures can be captured by unsupervised abnormal detection algorithms. [186, 187] detected abnormal graphs by monitoring the degree distribution of graphs. [29] used Self-organizing Maps (SOM) algorithm to filter large normal clusters, for reducing the search space of the botnet. More structural statistical features were gathered in [150]. It used 7 features to perform abnormal detection using one-class SVM and other abnormal detection algorithms. Graph edit distance was applied to model the traffic structural difference in [90], and the graph with a large average distance will be classified as a botnet. Unlike previous methods, they detected botnets at the graph level, as the graph is defined as a small portion (e.g., a traffic session) of the whole botnet.

Compared with the above methods which only utilize structural features, combining traffic flow features into botnet topology analysis will be more effective and robust. The ways to combine these features lie in two categories. On one hand, in a network traffic graph, nodes often represent hosts which have rich flow-based attributes, e.g., IP, port, and duration, thus both flow features and structural features can be incorporated. On the other hand, we can train the flow-based model and structure-based model separately and then integrate them to make the final decision. Among the first approaches, attribute-based statistical models [90, 150] extracted multiple flow-based and structural features from the host graph, altogether feeding into clustering/classification modules. In [237], multiple meta-paths and meta-graphs were first designed to capture the similarity between hosts. Then classical homogeneous embedding [86] was applied to obtain the final node representation to perform classification. As for the second approach, i.e., ensemble methods, the common way is modeling the flow-based and structure-based features independently. [190] found that the c-flow sequences of bots have a relatively stable length, while normal c-flow often mutates during passing. Thus a stability-based analysis is added to filter bot c-flows. The proposed model BotMark analyzes traffic flow and structural features in an ensemble way. Extensive experiments proved this model can considerably reduce the false positive rate.

Current graph-based solutions for botnet detection are mainly based on statistical features. With the increasing complexity of botnets (e.g., encrypted communications and self-destruction mechanisms), statistical features are hard to be constructed. Therefore, it is expected that more deep embedding methods are applied. Meanwhile, more botnets with novel types are emerging, while most existing methods are only suitable for specific known kinds. Zero-day botnet detection and adaptable graph-based models should be further explored in the future.

*5.3.2 Malicious Domain.* Blacklist is a basic and efficient method to filter malicious domains. However, criminals use Domain Generation Algorithms (DGA) to generate a large number of short-time domains which are hard to be all blocked by the blacklist. Even worse, with the development of domain fluxing technologies and the dynamic-changed IP addresses with DNS-based FastFlux tools [94], the blacklist mechanism is far from enough. Researchers then expect automatically detect malicious domains using learning-based methods, which can be divided into two categories: content-based and traffic-based. Content-based methods extract character features of the domain name while traffic-based methods extract static and dynamic flow features from DNS traffic such as Time To Live (TTL) values and reverse DNS query results [11]. These methods are effective but easy to be evaded by advanced attacks. The reason is that existing methods only consider the local features of individual domains, but ignore the relationship between domains, thus making the model not robust. Graph-based methods capture the global associations between domains and mine the deep semantic features between domains and other network components. The typical workflow of domain association mining is depicted in Fig. 9 and detailed solutions for mining these associations are introduced as follows.

**Domain Association Mining**. The insight behind domain association mining is that malicious domains tend to have similar behaviors. Therefore, the main challenge lies in two folds. (1) How to model the associations of domains. (2) How to use these associations to detect malicious domains. To model the associations of domains, a straightforward idea is that domains sharing the same IP address are likely similar. This simple assumption motivates researchers to construct *domain resolution graph* to capture the mapping relation between IP and domain [64, 84, 174]. [117] argued that this assumption does not always hold, and two domains are correlated only if they share at least one dedicated IP, or share more than one public IP from different hosting providers. [94] further constructed two other bipartite graph, namely *domain-host graph* and *domain-time graph*. The observations are that if two domains are queried by the same host, they tend to be strongly associated; Many domains show strong temporal correlations. In order to model explicit relationships between domains, [64, 84, 94, 117] exported domain similarity graph from domain resolution graph by one-mode projection. More complex associations between domains are modeled by the traffic activity graph in [164]. As shown in Figure 9, the intuitions of constructing this graph are (1) Domain character distribution. (2) The victims infected by the same crime tend to request the same domains. (3) The resource reuse of malicious domains.

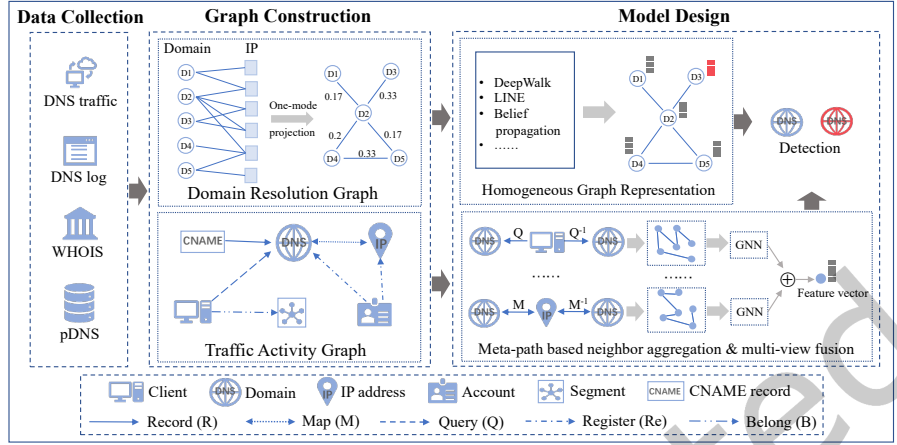

Fig. 9. The typical workflow of domain association mining in malicious domain detection. The data sources include DNS logs and traffic from the resolver in a LAN, passive DNS (pDNS), and WHOIS datasets from public services [164]. Two main graphs are constructed to perform domain association mining: (1) Domain resolution graph. It first performs one-mode projection to deduce domain similarity graph, then used homogeneous graph representation methods to perform node-level classification. (2) Traffic activity graph. It uses meta-path based neighbor aggregation and multi-view fusion to obtain feature vectors.

After constructing graphs to capture associations between domains, many graph mining methods are used to detect malicious domains. (1) Structural statistical features. [117, 174] directly ran the belief propagation algorithm on the domain resolution graph and propagate label information to get the malicious probabilities of unknown domains. [75] extracted 12 behavior features and used the random forest algorithm as the classifier. Given a set of known malicious domain nodes (seeds), [84] used path-based inference to calculate the maliciousness of unknown nodes based on the distance between these nodes and seeds. (2) Homogeneous embedding. LINE and DeepWalk were applied on the domain similarity graph by [94] and [64] respectively to obtain a richer representation of domains. (3) Heterogeneous embedding. To capture more domain semantic associations, as depicted in Figure. 9, [164] designed multiple meta-paths based on constructed traffic activity graphs. It aggregated neighbors that are sampled by meta-path guided random walk and utilized GraphSAGE to achieve inductive learning. In [165], the attention mechanism was further applied to fuse different semantic information.

Graph mining technologies have been widely used in malicious domain detection. Heterogeneous graph embeddings are proven to be the most effective and robust methods since they incorporate more domain features and behaviors. However, there are also some concerns that should be further explored. The first is DNS with stronger encryption protocols. Existing methods assume all the DNS traffic is available, but there already have some encrypted protocols (e.g., DNS over Transport Layer Security (TLS) (RFC 7858)) used in DNS traffic [164]. Designing graph-based methods for encrypted DNS is a challenging task. Second, lacking graph-based benchmark datasets. Engaging in collecting and releasing benchmark datasets is very important to advance the field.

*5.3.3 Intrusion Detection.* According to the technologies that Intrusion Detection Systems (IDS) used, we can divide IDS into misuse-based and anomaly-based. Misuse-based methods (also called signature-based methods) are rule-based methods, which use pre-defined signatures of malicious activities to identify intrusion behaviors, thus failing to detect zero-day attacks and needing manually update databases. Traditional anomaly-based methods can detect zero-day attacks but with a high false positive rate since all out-of-distribution behaviors are categorized as malicious. The graph-based intrusion detection methods are almost anomaly-based but include more graph modeling of networks and systems, which effectively decrease the false positive rate. We classify graph-based intrusion detection methods according to application scenarios, including alarm correlation mining, IDS components similarity mining, and IDS dynamic analysis. The typical workflow of these methods is depicted in Fig. 10 and details of them are introduced below.

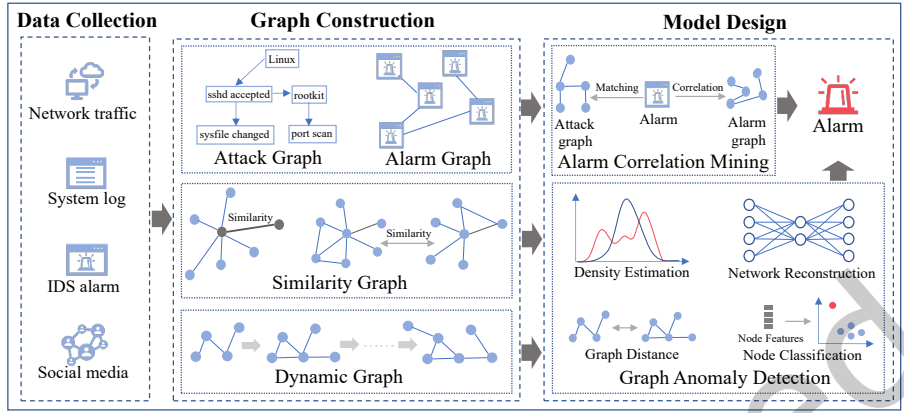

Fig. 10. The typical workflow of IDS using graph mining techniques. It improves the detection performance (alarm quality) in three ways: (1) Attack graphs guide the generation of alarms and alarm graphs capture the correlations among different alarms. (2) Similarity graphs are constructed to model the attribute and structural similarities among IDS components. (3) Dynamic graphs are constructed to monitor the chronological events in IDS. Based on similarity and dynamic graphs, multiple graph anomaly detection methods (e.g., network reconstruction) are applied to detect intrusions.

**Alarm Correlation Mining**. IDS generates massive alarms every day, coupled with false alarms. Alarm correlation mining aims to reduce the false positive rate of alarms and optimize the alarming quality by capturing the correlations between alarms or system behaviors. Attack graphs show all attack paths in the network that can be discovered by the defender, hence becoming a potent tool to guide the alarm correlation process. These methods are mostly based on statistical features. [231] used frequent patterns mining on multi-source logs to construct attack pattern graphs. Alarms will be triggered only if the suspicious attack matches the attack graph. Besides, by introducing the attack graph, the attack process can be restored for further analysis. Attack graphs can also determine the priority of alarms or find new alarm correlations [139]. [122] directly modeled the correlations between alarms in an alert graph, where nodes denote different properties of alarms such as file path and process, and edges denote co-occurrence weights. The role dynamic algorithm is applied to monitor the role distributions of nodes and determine whether to generate alarms.

**IDS Components Similarity Mining**. The goal of IDS components similarity mining is to use graphs to capture the similarities between IDS components. One example is the insider threat detection task. [45] modeled insider threat detection as a node classification task, which uses a user-device bipartite graph to represent the user's logging behaviors. Behavior patterns of malicious users are captured by structural features of the user's k-order neighbors. A similar idea was also used in [13]. The above methods only model the structural similarity but ignore the attribute similarity. In [142], to detect intrusions in cyber-physical systems, a KNN graph of sensors is constructed and the weights of the edges incorporate both sensors' structure similarity (geometric distance) and attribute similarity (measurement difference). It assumed the measurements follow Gaussian Markov Random Field (GMRF) distribution, thereby the out-of-distribution graphs are identified as abnormalities.

**IDS Dynamic Analysis**. The chronological events in IDS can be modeled as graphs. Therefore, the IDS dynamic analysis can be abstracted as dynamic graph anomaly detection [4], which aims to identify anomaly graphs or nodes in a graph sequence. In terms of outsider threat detection, the basic flow of graph-based solutions is discovering abnormal structural patterns through monitoring the dynamic TAGs in real time. Existing methods mainly utilized statistical features. [33] decomposed the host graph into subgraphs that correspond to a session. They give each subgraph an abnormal probability score by comparing the structural differences between newly traffic graph and the historical graphs. [92] conducted both static and dynamic analyses of host graphs. Static features (e.g., node degree and the entropy of the degree distribution) model the abnormality of the individual structure. Graph edit distance is used to model the dynamic structural abnormality. [77] reconstructed the adjacency matrix through a multi-layer autoencoder and identified abnormal traffic based on the reconstruction error. In [78], the TAGs were further enriched by introducing edge attributes and the abnormal scores are obtained by analyzing these attributes.

As for insider threat detection, [129] constructed a sequence of device-login graphs to model user daily remote logging behaviors. Given a sequence of historical device-login graphs, [129] used the role extraction algorithm to obtain the role vector of each node, then detected abnormal login nodes based on reconstruction losses of role vectors. In the Unmanned Aerial Vehicle (UAV) network, [27] constructed a UAV graph to model the communications between UAVs. The spatial relationships of nodes are captured by homogeneous embeddings. It used an SVM classifier to perform node-level anomaly detection. Besides these node-level anomaly detections, graph-level anomaly detection has been addressed in [85]. To maintain incrementally graph embedding and detect anomalous graphs in real-time, the representation of the graph was based on a subgraph structure called label structure, and graph edit distance was adopted to detect abnormalities.

Although these graph-based intrusion detection methods have achieved desired objectives in network management and protection, none of them proposed a unified graph-based framework for general IDS, which is imperative in the current complex and diverse network environment. Besides, detecting abnormal traffic is an essential but challenging task for IDS since network attacks such as Advanced Persistent Threat (APT) attacks and spam, the operations of network administrators, and a short-time increased page view can both bring on abnormal traffic. Thus, robust IDS should be paid more attention to. Also, time & space complexity are significant metrics to consider in graph-based IDS, which only be analyzed by several works [85, 142].

*5.3.4 Summary.* Graph-based traffic flow analysis considers both flow attributes and structures, which is more effective and robust compared with solely flow-based methods. Botnet topology analysis and domain association mining are two typical applications of graph-based traffic flow analysis. Intrusion detection uses both traffic flow analysis and internal components relation mining to ensure a safe and stable network environment. In fact, abnormal traffic is most likely a signal of serious cyber attacks, such as botnets, port scanning, and network worms. Thus the ability to detect abnormal ones has become indispensable for IDS. Noting that traffic classification is also a fundamental task for network security, which aims to categorize traffic flows based on different standards (e.g., protocols and applications) [137, 153]. With more fine-grained traffic flow representations, these crucial tasks in network security are promising to achieve higher performance. Besides, current network management systems are mostly based on rule guidance or statistical features, which is far from enough in an environment of low fault tolerance. Advanced graph embedding methods with scalability and interpretability are desired to integrate into these systems.

Table 6. Typical graph approach used in system security.

| Task | Graph Type | Graph Approach | Task Level | Paper |
|---|---|---|---|---|
| Malware | Function call | Structural (statistical feature) | Graph | [71, 210] |
| | | Attributed (statistical feature) | | [49] |
| | | | Subgraph | [38] |
| | | Str+Sha+Hom (graph embedding) | | [39] |
| | | Str+Dee+Hom (graph embedding) | Graph | [125] |
| | | Att+Dee+Hom (graph embedding) | | [16] |
| | File dependency | Structural (statistical feature) | Node | [220] |
| | | | Graph | [204] |
| | | Attributed (statistical feature) | | [68] |
| | | Str+Sha+Het (graph embedding) | | [40] |
| | | Att+Dee+Het (graph embedding) | Node | [188] |
| | | Str+Dee+Het (graph embedding) | | [218] |
| | File distribution | Structural (statistical feature) | | [73, 170] |
| | | Attributed (statistical feature) | | [5, 89] |
| | User interface | Structural (statistical feature) | Graph | [23] |
| System vulnerability | Vulnerability dependency | Structural (statistical feature) | Node | [136] |
| | Code property | | Graph | [184] |
| | | Attributed (statistical feature) | | [43, 208] |
| | | | | [35] |
| | | Att+Dee+Het (graph embedding) | | [183, 245] |
| Blockchain | Transaction | Structural (statistical feature) | Node | [24, 25, 44, 127] |
| | | Att+Sha+Hom (graph embedding) | | [201] |
| | | Att+Dee+Hom (graph embedding) | Subgraph | [6, 197] |
| | | | | [152] |
| | | Att+Dee+Het (graph embedding) | | [242] |
| | | | Node | [105] |
| | Code property | Str+Dee+Hom (graph embedding) | Graph | [72] |
| | | Att+Dee+Het (graph embedding) | | [108, 109, 246] |

## 5.4 System Security

Traditional methods for protecting system security ignore modeling the potential relationships of system components and thus may obtain unsatisfied results. For example, in malware detection, an unknown file that always co-occurrences with many Trojans possibly a malicious Trojan-downloader [40]. Traditional methods only utilize file contents thus being unable to identify this kind of malware. These limitations can be improved by graph-based methods. Therefore, in this section, we review existing graph-based malware and vulnerability detection, as well as the widely concerned blockchain system security solutions. We list related papers and corresponding graph approaches in Table 6 and will describe them in detail as follows.

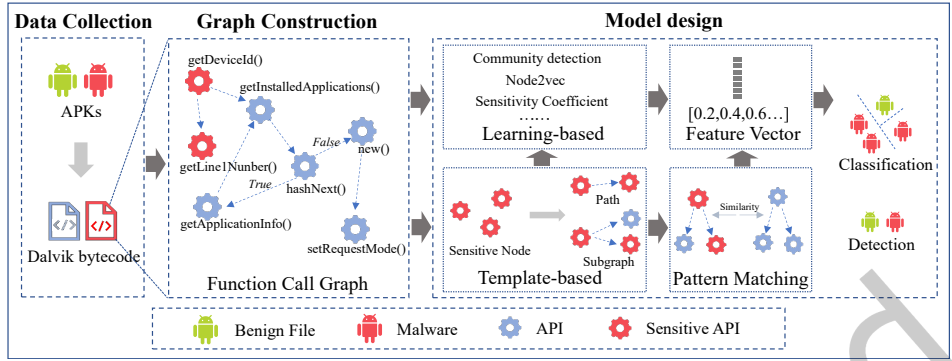

Fig. 11. The typical workflow of file-level analysis in malware detection. First, the Dalvik bytecode is obtained from the Android package (APK) through disassemble tools. Then the function call relations in Dalvik bytecode are captured by FCG. Based on FCG, template-based methods first construct sensitive templates (e.g., sensitive nodes, paths, and subgraphs), and then perform pattern matching to obtain feature vectors, while learning-based methods directly learn feature vectors through graph mining algorithms (e.g., Node2vec). The feature vectors are used for malware detection or family classification.

*5.4.1 Malware.* Using graph mining technologies, different granularities of malware structural information can be captured. Based on this partition, we divide existing graph-based malware analysis methods into four categories, including function-level, file-level, system-level, and network-level analysis. From the function level, relations between different blocks within a function can be captured by control flow graphs; from the file level, there exist multiple functions in code files and the inter-call relations between functions can be captured by the function call graphs; from the system level, the relations between code files and other system components (e.g., processes) can be captured by file dependency graphs; from the network-level, the relations of distributed files among network can be captured by file distribution graphs.

**Function-level Analysis**. Control Flow Graph (CFG) is widely used in function-level code analysis such as binary code similarity searching tasks [207, 222]. As a basic process of malware family classification, code similarity detection aims to find similar codes in the database or other platforms. [207] encoded CFGs of codes from different platforms by homogeneous embeddings and then used a siamese architecture to detect whether they are similar. [222] further considered the semantic and order information of blocks. CFG-based code representation is a fine-grained method for function-level tasks but ignores the relationship between functions. In addition, it is time-consuming to extract and analyze CFGs [96].

**File-level Analysis**. The goal of file-level malware analysis is to model the function (API) call relation or function dependency relation within a file (e.g., Android apps). The typical workflow is shown in Fig. 11. We can see that there exist two main approaches for file-level malware analysis, i.e., template-based methods and learning-based methods. Template-based methods first define some patterns of malware and then perform pattern matching between unknown files and known patterns. This process mainly utilizes structural statistical features. [71] first defined *Approximating Graph Edit Distance* for two FCGs to accelerate malware similarity searching. [210] proposed two-layer FCG where the upper layer models the interactions among android main components such as broadcast receiver and the lower layer models the API call relations within an upper component. To reduce the complexity of pattern matching, it only extracted call patterns of sensitive nodes. Inspired by social network analysis, [37] used community detection methods to extract sensitive subgraphs and performed clustering in these subgraphs to obtain structural patterns of the malware families. Template-based methods are simple and convenient but easy to be evaded and fail to detect zero-day malware.

Different from template-based methods, learning-based methods perform pattern learning and detection/classification simultaneously. Some sensitive structural statistical features were proposed in [38], such as the number of sensitive motif instances and total sensitive distance. [49] incorporated attributed statistical features (operations in the function) into the function call graph and used SVM with a neighborhood hash graph kernel to perform graph classification. It also assigned every subgraph a learnable weight to obtain interpretable results. To obtain higher-level semantics between API calls, [68] further proposed multiple meta-paths (e.g., API-Package-API) to formulate a similarity measure over Android apps. Structural embedding technologies such as node2vec and struct2vec are also used to represent the CFGs [39, 125]. The dense vector representation of CFG reduces time complexity by a big margin. Recently, [16, 96] used GNNs to incorporate CFG structure features as well as function attributes (i.e., function name), which makes the representation of CFG more effective and robust.

**System-level Analysis**. System-level malware analysis focuses on modeling file relations and the dependencies between files and other system components. The component interactions within a system are important behavior features to identify malware, especially for PC-based malware analysis. The file relation was explicitly modeled in [220]. It constructed a file co-occurrence graph where Jaccard similarity is used as edge weight. The file dependencies among system components are often modeled by the file dependency graph. Based on this kind of heterogeneous graph, traditional template-based methods are used to perform PC-based malware analysis [204]. Besides, many advanced graph embedding models are also widely applied. To consider semantic correlations among API calls and system entities, [40] designed multiple meta-paths for PC-based malware detection tasks, and then performed meta-path guided random walk to obtain node embeddings. In terms of unknown malware detection, [188] used GNNs and multi-view fusion technologies to obtain file embeddings. The maliciousness of an unknown file is determined by the similarity between its embedding and known benign files. Furthermore, [218] used GNNs and the representations of k-order neighbors to update node embeddings for inductive learning.

**Network-level Analysis**. The above methods achieved desirable performance by inspecting program contents (e.g., function and package). However, with the development of repackaging and obfuscation technologies, content-based methods have shown great limitations in identifying malware. As a result, researchers begin to study content-agnostic methods for malware analysis. Some useful characteristics are found in malware distribution networks. First, malware often maintains a relatively close distance from each other, for example, the maliciousness of files within a website may be the same. Second, the patterns between benign and malicious file download behavior in the network are distinguishable. These file distribution and downloading characteristics in the network motivate network-level malware analysis. Existing methods are mainly based on statistical features. For modeling distribution characteristics, the file-website bipartite graph was constructed in [170, 178], and belief propagation was applied to perform malware detection. [5, 73] considered more network entities and constructed the graph where nodes include file, IP, URL, etc. Based on this graph, [73] calculated the shortest path length between unknown files and malware, [5] incorporated both content-agnostic attributed feature and network topological information for malware detection. For modeling downloading characteristics, [89] constructed a file downloader graph and summarized discriminating features (e.g., domain names, download time, file behavior, etc.) for identifying malware.

Malware has been around since the appearance of computers. Static analysis has been extensively applied by modeling its inherent graph structures (e.g., CFG and FCG). Generally, the methods with fine-grained analysis consider richer semantic information but suffer higher computation costs. The complexity of graph-based solutions should be carefully considered to accommodate the rapid evolution of malware families. Meanwhile, the emerging various adversarial techniques (e.g., obfuscation and encryption) may make the malware not distinguishable by the static analysis but they are hard to evade the dynamic analysis. However, dynamic analysis is more time-consuming. Therefore, it's challenging to conduct dynamic analysis with high efficiency by means of graph mining techniques.

*5.4.2 System Vulnerability.* Traditional vulnerability detection technologies include static and dynamic approaches based on whether needing to run the systems. Static approaches focus on extracting features such as lexical, control flow, and data flow from source code to detect vulnerabilities, but with high false positives and low accuracy. Dynamic analysis such as fuzz testing and taint analysis executes the code in real systems or emulators and detects vulnerabilities by monitoring the running states, however, they suffer scalability issues. Nowadays, static analysis remains the main approach to detect vulnerabilities [99]. The major limitation of traditional static analysis is that they only model the sequential features of codes [245]. In fact, source codes include rich structural and semantic information, which can be depicted by graphs such as Abstract Syntax Tree (AST) and Program Dependence Graph (PDG). As a result, the sequential-based static analysis is far from sufficient to model various vulnerabilities. Therefore, graph-based methods have emerged in recent years. In this section, we summarize existing graph-based static vulnerability mining works, which can be divided into two aspects: inter-relation mining and intra-relation mining.

**Vulnerability Inter-relation Mining**. When evaluating vulnerability threats, traditional methods analyze vulnerabilities individually, ignoring the correlation between vulnerabilities. For example, attackers often utilize multiple vulnerabilities to conduct multi-step attacks from different paths. The key vulnerabilities should be identified and given high-risk value to further protect all subsequent vulnerabilities even the whole system. Therefore, when assessing the threat of vulnerabilities, we must consider the inter-relations between vulnerabilities.

The inter-relations between vulnerabilities can be modeled as Vulnerability Dependency Graph (VDG), which is induced from attack graphs. [136] proposed a vulnerability correlation hazard assessment method based on VDG and risk matrix. It not only utilized the features of vulnerabilities but also considered the correlation between vulnerabilities in the VDG to assess the threat degree of vulnerabilities. The overall system security can be assessed by aggregating the risks of vulnerabilities. Common Vulnerability Scoring System (CVSS) is a standard to measure the severity of vulnerabilities. However, CVSS also fails to model inter-relations between vulnerabilities. To solve this issue, [184] proposed an improved system security metrics algorithm based on VDG and CVSS. It aggregated the dependency relation of vulnerabilities into the basic measurement algorithm of CVSS and obtained scores in terms of the probability and impact of the system being attacked.

**Vulnerability Intra-relation Mining**. There are three typical graphs to represent programs, that is, Control Flow Graph (CFG), Abstract Syntax Tree (AST), and Program Dependence Graphs (PDG). Intra-relation mining aims to model the spatial structure as well as other attributes of programs based on these graphs. Early methods are mostly based on attributed statistical features. For example, [43] used CFG to represent programs and convert raw features (CFGs) into high-level numeric vectors to perform scalable vulnerability searching. Code Property Graph (CPG) [208] merged the above three graphs into a unified graph, which includes richer information about programs. [208] defined the traversal algorithm on the CPG. This traverse can characterize a large part of code vulnerabilities and discover new vulnerabilities. [221] used graph traversal on CPG to execute taint analysis and detect Server-Side Request Forgery (SSRF) vulnerabilities based on the constraint solving method. CPG contains rich structural features of programs but ignores the code sequential features. Many augmented CPGs were proposed by adding additional edges to model sequential features and other information of codes [245]. Based on augmented CPGs, the advanced methods mainly utilized heterogeneous graph embeddings to perform vulnerability detection [245].

Similar to malware detection, system vulnerability detection is also based on code analysis, especially focusing on function- or file-level analysis. The main difference between these two tasks is that vulnerability detection pays more attention to the semantics of codes since it focuses on detecting programming defects. Therefore, besides some insights present in Section 5.4.1, semantic-preserving vulnerability detection is also a key point when applying graph mining techniques [104]. Existing works have already constructed some semantic-preserving graphs into code analysis, such as AST and CPG [208, 245], and more related graph types and methods (e.g., meta-paths for specific semantic information ) are expected to be present. The more fine-grained vulnerability detection is another research trend. From application level, file level, to function level, it's desired that graph mining technologies can locate vulnerabilities more accurately, e.g., segment level.

*5.4.3 Blockchain Security.* To tackle blockchain security risks, many efforts have been made by the blockchain community and researchers. Traditional network attacks are common threats to cyber systems and many tailored intrusion detection models are proposed [157]. The rule-based methods are dominant in improving the defects of blockchain design. For example, studies design new transaction rules to prevent the typical 51% Attack [149]. As for the criminal activities in the blockchain, traditional methods mainly focus on analyzing important properties of the activity (e.g., the amount of transferring money). Besides the above efforts, considering the blockchain is a peer-to-peer network in nature, graph mining technologies have unique advantages in portraying the various structural patterns of blockchain. In this section, we introduce two main threats to blockchain security, namely smart contract vulnerability and criminal activities in the blockchain, as well as corresponding graph-based solutions.

**Smart Contract Vulnerability**. To detect smart contract vulnerability, like vulnerability intra-relation mining that we discussed in Section 5.4.2, current graph-based approaches mainly model the smart contract as a code graph. Structure-based graph embedding was used in [72]. It first constructed CFGs by simulating the bytecode execution of smart contracts, then used graph2vec [119] to obtain code representations. Heterogeneous GNNs are further used to model complex semantics of code graphs [108, 109, 246]. They model the source code as a *contract graph*, whose nodes denote invocations or variables and edges denote multiple semantic relations (e.g., control flow and data flow). Based on the contract graph, [246] proposed a temporal message propagation network (TMP) to capture temporal relations of different program elements. [108, 109] further integrated expert vulnerability patterns (e.g., an invocation of *call.value* is relevant with reentrancy vulnerability) to improve accuracy.

**Criminal Activities in the Blockchain**. By modeling activities of blockchain in a transaction graph, current graph-based solutions focus on identifying anomalies of accounts, transactions, and smart contracts in the blockchain system. Another relevant task is account identity inference (also called de-anonymization), which aims to infer the possible identity of accounts in blockchain, such as exchanges, phishing accounts, miners, etc [242]. We will introduce these solutions according to the format of constructed transaction graphs.

Based on the transaction graph whose nodes denote accounts and edges denote transactions, [44, 127] used structural statistical features (e.g., PageRank and clustering coefficient) to identify suspicious accounts. [25] further adopt temporal structural analysis to prove the market manipulation of Bitcoin. [201] added timestamp and amount to the transaction edges and incorporated this semantic information to learn node embeddings. [197] first defined the trust values of the nodes and then propagated trust values through GNNs. To perform account identity inference, [152] extracted account subgraphs (k-hop neighbors) and then used GNN to obtain account representations. Another kind of transaction graph whose nodes denote transactions and edges denote payment flows is used in abnormal transaction detection [6, 127]. Besides, some works incorporated more blockchain entities and complex relations into the transaction graph [24, 105, 132]. Specifically, [24] constructed three kinds of graphs, which model money (Ether) transactions, smart contract creation, and invocation respectively. Based on these graphs, structural statistical features were proposed to detect anomalies and identify accounts in Ethereum. [105] modeled these blockchain entities in a unified heterogeneous graph and used a Heterogeneous Graph Transformer Network (HGTN) to detect abnormal smart contracts.

As a newly emerged cyber system, blockchain poses great challenges to current system security. The smart contract vulnerability and criminal activities have been widely addressed by advanced graph embedding techniques. However, enhancing the privacy-preserving ability is still a pain point for building blockchain systems. With the ever-growing Decentralized Applications (DApps) and online transactions, privacy-preserving graph algorithms are urgently needed, with some typical privacy techniques such as mixing and homomorphic encryption incorporated. Meanwhile, it's expected to achieve a well-balanced between efficiency, effectiveness, and privacy-preserving of the algorithms.

*5.4.4 Summary.* The three typical system security tasks introduced above are mainly based on graph-based code analysis. Therefore, a well-designed code representation algorithm is significant for these downstream tasks. Considering the rapid development of various advanced attacks (e.g., new malware families and vulnerabilities), robust code representation is the current research hotspot. A few works have designed attack&defence means for graph-based code representation model [2, 67]. Also, more graph types should be explored to improve the effectiveness and robustness of system security solutions. For example, [23] constructed User Interface Graph (UIG) to discriminate between benign and malicious codes without inspecting their contents. Besides blockchain systems, with the explosive growth of IoT devices, graph-based solution for the security of IoT system is also a mainstream research direction [219].

# 6 OPEN DATASETS AND TOOLKITS

In this section, we summarize existing open datasets and toolkits for graph-based cybersecurity solutions.

## 6.1 Open Datasets

Over the years, many graph-based cybersecurity datasets have been released to the public for further research. We summarize these resources in Table 7, including URL, publish year, label, and related papers. We organize these datasets according to the cybersecurity task taxonomy in Section 2.1. Note that due to privacy, there are few datasets in some tasks (e.g., financial fraud and underground market). Here we introduce representative datasets among them.

**Elliptic**. This Bitcoin transaction data is released by Elliptic company for anti-money laundering/blockchain security tasks with over 200K Bitcoin transactions (nodes), 234K directed payment flows (edges), and 166 node features. 4,545 nodes (2%) are illicit and 42,019 nodes (21%) are licit. The remains are not labeled. It uses a heuristics-based reasoning method for labeling. For example, illicit transactions tend to execute transactions with a lower number of inputs [199].

**WEBSPAM-UK2007**. This dataset is provided for research on web spam detection tasks. This is a large collection of annotated spam/nonspam hosts labeled by volunteers. Besides, the dataset also contains hyperlinks and HTML page content. Within the labeled dataset 5.19% was 'spam' and 88.33% was 'non-spam'. The rest was labeled 'undecided'.

**Twitter15&Twitter16**. These two datasets contain a collection of source tweets with their corresponding retweets and replies in 2015 and 2016. There are four different labels, False Rumor (FR), Non-Rumor (NR), Unverified (UR), and True Rumor (TR). The source tweets are annotated by referring to the labels of the events they are from.

**LIAR**. LIAR is a large fake news detection dataset that includes 12,836 short statements with subject, context/venue, speaker, state, party, and prior history. It is collected from POLITIFACT.COM's API[4] with a grounded and natural context, such as political debate, TV ads. LIAR considers six fine-grained labels: pants-fire, false, barely-true, half-true, mostly-true, and true. These labels are evaluated by POLITIFACT.COM editors for their truthfulness.

**CTU-13**. CTU-13 is a popular public benchmark dataset of botnet traffic that is from 13 scenarios (e.g., ClickFraud, PortScan). For all the scenarios, CTU-13 dataset converts the captured pcap files to NetFlows and releases the processed flows. There are three types in the label set, namely background, botnet, and normal. The normal labels are assigned by some filters and a botnet label is assigned if the traffic comes from or to know infected IP addresses [47].

---
[4]https://www.politifact.com

Table 7. Typical public graph-based cybersecurity datasets. TS: Transaction Security; CS: Cognition Security; NS: Network Security; SS: System Security.

| | Task | Dataset | URL | Year | Label | Paper |
|---|---|---|---|---|---|---|
| TS | Financial fraud | Elliptic | https://www.elliptic.co/blog/elliptic-dataset-cryptocurrency-financial-crime | 2019 | Licit/illicit | [199] |
| | | Czech Financial 1999 | https://data.world/lpetrocelli/czech-financial-dataset-real-anonymized-transactions/ | 1999 | None | [98] |
| | Underground market | D-GEF | https://github.com/HongyiZhu/D-GEF | 2020 | Attack type | [143] |
| | | CrimeBB | https://www.cambridgecybercrime.uk/datasets.html | - | None | [124] |
| CS | Web spam | WEBSPAM-UK2007 | https://chato.cl/webspam/datasets/uk2007/ | 2007 | Spam/non-spam | [144] |
| | | WEBSPAM-UK2006 | https://chato.cl/webspam/datasets/uk2006/ | 2006 | spam/non-spam | [1] |
| | Fake news | Twitter15/16 | https://www.dropbox.com/s/7ewzdrbelpmrnxu/rumdetect2017.zip?dl=0 | 2017 | Fake/non-fake/uncertain/true | [10, 110, 223] |
| | | News Aggregator | https://archive.ics.uci.edu/ml/datasets/News+Aggregator | 2018 | None | [202] |
| | | PHEME | https://github.com/azubiaga/pheme-twitter-conversation-collection | 2017 | Rumour/Non-rumour | [120, 193, 213] |
| | | FakeNewsNet | https://github.com/KaiDMML/FakeNewsNet | 2018 | Fake/real | [46, 120, 243] |
| | | LIAR | http://www.cs.ucsb.edu/~william/data/liar_dataset.zip | 2017 | False/true/half-true, etc. | [211] |
| | | weibo | http://alt.qcri.org/~wgao/data/rumdect.zip | 2016 | Rumour/non-rumour | [10, 193, 223] |
| | Review spam | Amazon review1 | http://liu.cs.uic.edu/download/data/ | 2006 | None | [195, 196] |
| | | Amazon review2 | http://jmcauley.ucsd.edu/data/amazon/ | 2014 | None | [121, 195] |
| | | YelpChi | http://shebuti.com/collective-opinion-spam-detection/ | 2013 | Spam/non-spam | [135, 195] |
| | | YelpNYC | http://shebuti.com/collective-opinion-spam-detection/ | 2015 | Spam/non-spam | [135, 195] |
| | | YelpZip | http://shebuti.com/collective-opinion-spam-detection/ | 2015 | Spam/non-spam | [135, 195] |
| | | op_spam | http://www.cs.cornell.edu/myleott/op_spam | 2011 | Truthful/deceptive truthful/deceptive | [121] |
| | | SMS/review | https://github.com/Giruvegan/stoneskipping/tree/master/dataset/review | 2019 | Spam/non-spam | [80] |
| | Fake account | Twitter10 | http://an.kaist.ac.kr/traces/WWW2010.html | 2010 | None | [145] |
| | | MIB | http://mib.projects.iit.cnr.it/dataset.html | 2015 | Fake/benign | [97] |
| | | Vendor-19 | https://botometer.osome.iu.edu | 2019 | Fake/benign | [214] |
| NS | BotNet | DDoS Attack 2007 | http://www.caida.org/data/passive/ddos-20070804dataset.xml | 2013 | None | [186, 187] |
| | | Twente traffic traces | http://eprints.eemcs.utwente.nl/17829/ | 2010 | Botnet/normal | [186] |
| | | CTU-13 | https://www.stratosphereips.org/datasets-ctu13 | 2014 | Background/botnet/normal | [29, 90, 187] |
| | | BotMark | http://infosec.bjtu.edu.cn/wangwei/page_id=85 | 2020 | Botnet/normal | [190] |
| | | CERT Insider Threat | https://www.cert.org/insider-threat/tools/ | 2016 | Malicious/benign | [45] |
| | IDS | NSL-KDD | https://www.unb.ca/cic/datasets/nsl.html | 2009 | Normal/DOS/Probe/R2L/U2R | [27] |
| | | CICIDS2017 | https://www.unb.ca/cic/datasets/ids-2017.html | 2017 | Attack Type | [27] |
| | | StreamSpot | https://github.com/sbustreamspot/sbustreamspot-data | 2016 | None | [85] |

Continued

| | | | | | | |
|---|---|---|---|---|---|---|
| SS | | USTC-TFC2016 | https://github.com/echowei/DeepTraffic/tree/master/1.malware_traffic_classification | 2017 | None | [163] |
| | | ISCX VPN-nonVPN | https://github.com/louiseviden/ns18 | 2016 | Application type | [163] |
| | Malware | Genome project | http://www.malgenomeproject.org | 2012 | Attack type | [96, 187] |
| | | drebin | https://www.sec.tu-bs.de/~danarp/drebin/download.html | 2014 | Malware/benign | [96, 125, 187] |
| | | FalDroid | https://github.com/xjtu1025/FalDroid | 2018 | Family | [96, 187] |
| | | androzoo | https://androzoo.uni.lu/lists | 2017 | Malware/benign | [16, 125] |
| | | malicia | http://malicia-project.com | 2013 | Family | [204] |
| | | Virusshare | https://virusshare.com | - | Malware | [37] |
| | System vulnerability | Devign | https://sites.google.com/view/devign | 2019 | Vulnerability type | [245] |
| | | Draper VDISC | https://osf.io/d45bw/ | 2018 | Vulnerability type | [42] |
| | | FUNDED | https://github.com/HuantWang/FUNDED_NISL/tree/main/FUNDED/data/data | 2020 | None | [183] |
| | | SARD | https://www.nist.gov/itl/ssd/software-quality-group/samate/software-assurance-reference-dataset-sard | - | Vulnerability type | [35] |

**NSL-KDD**. This dataset includes a wide variety of intrusions simulated in a military network environment. There are four main attack types in this dataset, namely DoS, probing, U2R, and R2L. Every sample has 38 numerical features with three content features. These features are mainly based on basic TCP connections and content features are collected by domain knowledge within a connection.

**Amazon Review**. This dataset focuses on review spam detection tasks and involves multiple reviews. Amazon review1 dataset contains information about reviewers from 1996 to 2006 and corresponding review text, ratings, products, etc. As the whole dataset is extremely large, many studies only extracted the book review data from the dataset. The Amazon review2 dataset contains product reviews (ratings, text, helpfulness votes), and metadata (descriptions, category information, etc) from Amazon from May 1996 to July 2014.

**SARD**. This dataset contains a growing collection of almost two hundred thousand programs with documented vulnerabilities. These vulnerabilities have different code forms (source code or binary code), different languages (C, Java, Python, etc.), and different types (buffer error, resource management error, injection, etc.), which cover over 150 classes of weaknesses. These vulnerabilities are collected in many ways, such as manual injection and static analysis.

**Drebin**. This labeled mobile malware dataset is an extension of the Genome project dataset and contains 179 different malware families. The samples were collected from August 2010 to October 2012. The VirusTotal service is used to determine whether an application is malicious or benign. Note that VirusTotal is also widely used in other malware dataset labeling [37].

## 6.2 Toolkits

Graph mining and data collection in cybersecurity solutions both benefit from public toolkits. However, few surveys have summarized these resources. To bring this gap, we present some typical graph mining and data collection toolkits.

*6.2.1 Graph Mining Toolkits.* Graph model implementation is a crucial process in cybersecurity solutions. To facilitate this process, many toolkits have been developed to allow researchers to easily conduct experiments and build applications. These toolkits provide standard training and evaluation for important graph-based tasks, including node classification, link prediction, graph classification, etc. Here we list typical ones of them for reference.

**PyG**. PyG is a machine learning library built upon PyTorch to easily implement Graph Neural Networks and related applications. It covers a wide range of state-of-the-art GNN architectures and training and scalability procedures, which can help users easily reproduce and design GNN experiments. The website of PyG can be found in https://www.pyg.org.

**CogDL**. CogDL is an extensive toolkit for deep learning on graphs. Most models in CogDL are developed based on PyTorch, with high efficiency and reproducibility. It provides easy-to-use APIs for running experiments and utilizes well-optimized operators to speed up training and save Graphics Processing Unit (GPU) memory of GNN models. The project can be found in https://github.com/thudm/cogdl.

**Deep Graph Library (DGL)**. DGL is a python package building on top of the current prevalent framework (Pytorch, MXNet, and Tensorflow). Besides homogeneous graphs, DGL also supports many heterogeneous graph models (e.g., HAN, Metapath2vec). This project can be found in https://github.com/dmlc/dgl.

**OpenHGNN**. This is an Open-source toolkit for Heterogeneous Graph Neural Networks (OpenHGNN) based on DGL. It provides easy-to-use interfaces for running experiments with many popular heterogeneous graph models, including RGCN, HAN, KGCN, HetGNN, GTN, etc. This project can be found in https://github.com/BUPT-GAMMA/OpenHGNN and the documentation is in https://openhgnn.readthedocs.io/en/latest/.

**Gamma Graph Library (GammaGL)**. GammaGL is an open-source graph learning library, which supports TensorFlow, PyTorch, PaddlePaddle, and MindSpore as the backends. Different from DGL, GammaGL's examples are implemented with the same code on different backends. It allows users to run the same code on different hardware and use a particular framework API based on preferences for different frameworks. This project can be found in https://git.openi.org.cn/GAMMALab/GammaGL or https://github.com/BUPT-GAMMA/GammaGL.

*6.2.2 Data Collection Toolkits.* Considering the rapid evolution of the cyberspace security situation, many datasets used in cybersecurity require updating frequently. Data collection is never a trivial process since the raw data in cybersecurity presents various forms (e.g., pcap) and should be processed in advance. Besides, positive samples (e.g., malicious traffic) are rare in the real world, making the data collection process more challenging. Thanks to public data collection tools, one can avoid collecting data from scratch. We summarize typical tools as follows.

**Twitter Search API**. It is an API that allows us to find and retrieve, engage with, or create a variety of different data sources including tweets, retweets, comments, etc, which is an important tool to collect fake news datasets, such as retrieving tweets of interest by giving certain queries. The API is in https://dev.twitter.com/rest/public/search.

**Snopes**. Snopes is a fact-checking site, which can be used to label rumors and non-rumors samples. The website is in http://www.snopes.com/.

**DNSDB Scout**. DNSDB is a database that stores and indexes passive DNS data provided by Farsight Security's Security Information Exchange and authoritative DNS data provided by various zone operators. The database also keeps first-seen and last-seen timestamps of domain-IP resolutions. The DNS data in this database can be obtained through the website's API: https://scout.dnsdb.info.

**Semantic**. Semantic is a Haskell library and command line tool for parsing, analyzing, and comparing source code. It can produce an Abstract Syntax Tree (AST) for each function of source code and support more than ten languages, including Python, Java, PHP, etc. This tool can be found in https://github.com/github/semantic.

**VirusTotal**. VirusTotal is a platform that analyzes suspicious files and URLs to detect types of malware and automatically shares them with the security community. It is commonly used to label samples as malicious or not. The website is in https://www.virustotal.com/en/.

**Cuckoo Sandbox**. Cuckoo Sandbox is an open-source dynamic malware analysis system. It spans four operating system platforms (Windows, Android, Linux, and Darwin) and supports multiple file forms (executables, office documents, pdf files, emails, etc). It can trace API calls and the general behavior of the file, which is very helpful for the dynamic analysis of malware. The URL of this tool is http://www.cuckoosandbox.org/.

**Wireshark**. Wireshark is widely used for analyzing network traffics. This tool can capture packets on the internet, along with various types of traffic features like IP, port, packet length, and protocol. These features can be further utilized for malicious traffic detection. One can download Wireshark at https://www.wireshark.org.

## 7 DISCUSSION AND FUTURE DIRECTIONS

Despite the wide application of graph mining technologies to cybersecurity, proposing an ideal graph-based cybersecurity solution is still challenging. The increasingly complex cyberspace environment and evolving criminal behaviors call for exploring new perspectives to design graph-based models. In this section, we give a discussion of existing graph-based cybersecurity solutions and explore some future research directions.

### 7.1 Graph Construction and Datasets

Cybersecurity tasks are highly data-driven. However, due to privacy as well as the difficulties to access real cyber environments, public graph-based benchmark datasets are extremely rare. Despite some efforts to carry out this work recently (e.g., labeled Bitcoin transactions datasets [199] and hacker posts datasets [143]), there is still a long way to go to create and release high-quality graph-based cybersecurity datasets. This issue makes researchers have to collect data from the real world. However, the sparseness, implicit relations, and numerous noises of real-world cyberspace data make graph construction a non-trivial task. Existing works mainly rely on domain knowledge and manual construction. Take malicious domain detection task as an example [165]. Mining implicit relations (e.g., segment and canonical name relations) for identifying malicious domains call for some domain knowledge (e.g., adjacent clients are prone to be infected by the same attacker and the properties of domains with the same canonical name are similar). Besides, the data collected from DNS traffic are full of noises (e.g., inactive clients and popular domains) and some graph pruning strategies should be applied. In fact, manually constructed graphs may not be optimal for specific cybersecurity tasks, thus how to automatically learn an optimal or better graph structure from cyberspace data is desired in the future.

### 7.2 Trustworthy Model and Robustness

Although existing graph-based cybersecurity solutions have achieved high performance, most of them lack consideration of model trustworthiness and robustness. In fact, a trustworthy and robust model is highly desired in the complex and adversarial cyberspace environment. Interpretability (i.e., the ability to interpret the decisions made by trained models) is significant for trustworthy models. For example, interpretability can enhance the trustiness of the public towards the fact-checking results of the news; In traffic analysis, interpretability can provide clear traits of malicious traffic for network managers. Several works have introduced the attention mechanism to analyze the importance among node neighbors [180, 240]. [180] selected important apps for default prediction based on attention value and finds that these apps are mainly financial apps, which is reasonable and yet enhances the reliance on proposed models. However, in the cybersecurity field, the existing interpreting models only focus on node classification tasks (e.g., default user detection), but fail to interpret the subgraph/graph-level tasks (e.g., malware detection). In addition, the importance of the sub-structure in the graph should also be explored.

Robustness is another desired property in today's graph-based cybersecurity solutions. Adversarial behaviors are ubiquitous in cyberspace, such as data perturbation [53] and camouflaged fraudsters [34]. Many existing models show weakness when suffering these adversarial behaviors, thus more robust models are urgently needed. A few works have considered this issue but have been limited to specific tasks [2, 67]. Meanwhile, yet maintaining good performance is still a challenge. Besides, existing graph-based models seldom consider the zero-day threats, which are more common in real-world environments [137]. Unsupervised methods (e.g., clustering) are expected to be incorporated to detect zero-day threats, hence improving the robustness of existing cybersecurity solutions.

## 7.3 Data Security and Privacy

The real-world datasets in cybersecurity often contain much private information, such as user accounts and passwords, making it hard for deep and comprehensive data sharing. Therefore, it is necessary to study the graph collaborative analysis architecture suitable for cybersecurity data, under the condition that sensitive parameters are not to be leaked. In recent years, many privacy-preserving collaborative graph mining techniques have emerged, which aim to perform secure data sharing among trusted members [63, 229]. Several works have also explored their application on some cybersecurity tasks. For example, [168] proposed a federated graph learning platform to share key information across institutions for efficiently detecting global money laundering activity. More data-sensitive security scenarios are desired to be explored. Besides, introducing privacy-preserving algorithms often causes additional computing costs and performance decay. Therefore, privacy-preserving graph mining while maintaining efficient computations and high performance is expected to be further researched in the future.

## 7.4 Large-scale Graph and Parallelization

The real-time interactions between cyberspace entities make the constructed graph dynamic and large-scale. For example, the online fraud detection system takes approximately 1 hour to score 200 million customers [70]; there are hundreds of parameters of network nodes and connections in the IoT scenarios [159]. This means that the graph model should be expressive enough to describe adaptive behavior and handle large-scale graphs. Many scalable graph algorithms have been proposed in recent years, including graph inductive learning [60], efficient sampling [22], graph distillation [232], etc. These algorithms have been successfully applied to several cybersecurity tasks (e.g., malicious domain detection [164] and fraud detection [199]). With respect to real-world deployments of large-scale graphs, distributed and parallel graph training strategies are proposed [112, 226, 238]. These strategies partition graph data into a cluster of machines, and train graph models in a parallel fashion. As an instance, the Ant Graph machine Learning system (AGL) [226] decomposed the original graph into pieces of subgraphs (i.e., K-hop neighborhood) for message passing, which is simply implemented by MapReduce [31] and utilized in financial risk control with high efficiency. With the ever-growing interactions of cyberspace entities, it's necessary to incorporate scalable graph models and large-scale deployment strategies into cybersecurity solutions.

## 7.5 Attacker Correlation Analysis and Discovery

At present, the application of graph mining techniques to cybersecurity mostly focuses on modeling the correlations between attack elements to perform attack detection or classification (e.g., API call relations within a file are captured by API call graphs). However, few researchers use the advantage of graph mining techniques in correlation analysis to correlate and find attackers (or attack organizations). The attack organization information, including its background, tools, and malicious samples, obtained in the process of digital forensics and tracing has become an important resource in the network attack defense. Meanwhile, this resource also reveals the correlations between organizations. For example, it's reported that there is a lot of code reuse among attack organizations in North Koreans, which indicates that groups with different skills and tools will execute their focused parts of cyber attacks while also working in parallel when collaborating on large attacks [140]. Therefore, using graph mining techniques to mine the deep correlations between attack organizations based on massive organization information and utilize these correlations to perform network attack traceability is a promising research point. From a more macroscopic point of view, it is expected that applying graph mining techniques to promote the association and coordination between defenders in the future.

## 8 CONCLUSIONS

In this survey, we conduct a comprehensive overview of the application of graph mining techniques to cybersecurity. We provide a new taxonomy of typical cybersecurity tasks based on applied graph mining techniques. We summarize typical graph mining techniques used in cybersecurity, including statistical features and graph embedding. we also present general steps for applying these techniques to cybersecurity. Then, we elaborate on these graph-based cybersecurity solutions. Besides, we summarize public cybersecurity datasets and toolkits for further research. Finally, we suggest five promising research directions.

## ACKNOWLEDGMENTS

This work was partially supported by the National Natural Science Foundation of China (No. U20B2045, 62192784, 62172052, 62002029, 61772082, 62172055) and BUPT Excellent Ph.D. Students Foundation (No. CX2021118).

## REFERENCES

[1] Jacob Abernethy, Olivier Chapelle, and Carlos Castillo. 2010. Graph regularization methods for web spam detection. *Machine Learning* 81, 2 (2010), 207–225.

[2] Ahmed Abusnaina, Aminollah Khormali, Hisham Alasmary, Jeman Park, Afsah Anwar, and Aziz Mohaisen. 2019. Adversarial Learning Attacks on Graph-based IoT Malware Detection Systems. In *ICDCS*. IEEE, 1296–1305.

[3] Leman Akoglu, Rishi Chandy, and Christos Faloutsos. 2013. Opinion Fraud Detection in Online Reviews by Network Effects. The AAAI Press.

[4] Leman Akoglu, Hanghang Tong, and Danai Koutra. 2015. Graph based anomaly detection and description: a survey. *Data Min. Knowl. Discov.* 29, 3 (2015), 626–688.

[5] Ibrahim Alabdulmohsin, YuFei Han, Yun Shen, and Xiangliang Zhang. 2016. Content-agnostic malware detection in heterogeneous malicious distribution graph. In *CIKM*.

[6] Ismail Alarab, Simant Prakoonwit, and Mohamed Ikbal Nacer. 2020. Competence of graph convolutional networks for anti-money laundering in bitcoin blockchain. In *Proceedings of the 2020 5th International Conference on Machine Learning Technologies*. 23–27.

[7] Hisham Alasmary, Aminollah Khormali, Afsah Anwar, Jeman Park, Jinchun Choi, Ahmed Abusnaina, Amro Awad, DaeHun Nyang, and Aziz Mohaisen. 2019. Analyzing and Detecting Emerging Internet of Things Malware: A Graph-Based Approach. *IEEE Internet Things J.* 6, 5 (2019), 8977–8988.

[8] Luca Becchetti, Carlos Castillo, Debora Donato, Stefano Leonardi, and Ricardo A Baeza-Yates. 2006. Link-based characterization and detection of web spam. In *AIRWeb*. 1–8.

[9] Alex Beutel, Wanhong Xu, Venkatesan Guruswami, Christopher Palow, and Christos Faloutsos. 2013. CopyCatch: stopping group attacks by spotting lockstep behavior in social networks. In *WWW*. International World Wide Web Conferences Steering Committee / ACM, 119–130.

[10] Tian Bian, Xi Xiao, Tingyang Xu, Peilin Zhao, Wenbing Huang, Yu Rong, and Junzhou Huang. 2020. Rumor detection on social media with bi-directional graph convolutional networks. In *AAAI*.

[11] Leyla Bilge, Engin Kirda, Christopher Kruegel, and Marco Balduzzi. 2011. EXPOSURE: Finding Malicious Domains Using Passive DNS Analysis.. In *NDSS*. 1–17.

[12] Yazan Boshmaf, Dionysios Logothetis, Georgos Siganos, Jorge Lería, José Lorenzo, Matei Ripeanu, and Konstantin Beznosov. 2015. Integro: Leveraging Victim Prediction for Robust Fake Account Detection in OSNs. In *NDSS*. The Internet Society.

[13] Oliver Brdiczka, Juan Liu, Bob Price, Jianqiang Shen, Akshay Patil, Richard Chow, Eugene Bart, and Nicolas Ducheneaut. 2012. Proactive Insider Threat Detection through Graph Learning and Psychological Context. In *IEEE Symposium on Security and Privacy Workshops*. IEEE Computer Society, 142–149.

[14] Adam Breuer, Roee Eilat, and Udi Weinsberg. 2020. Friend or Faux: Graph-Based Early Detection of Fake Accounts on Social Networks. In *WWW*. ACM / IW3C2, 1287–1297.

[15] Hongyun Cai, Vincent W. Zheng, and Kevin Chen-Chuan Chang. 2018. A Comprehensive Survey of Graph Embedding: Problems, Techniques, and Applications. *IEEE Trans. Knowl. Data Eng.* 30, 9 (2018), 1616–1637.

[16] Minghui Cai, Yuan Jiang, Cuiying Gao, Heng Li, and Wei Yuan. 2021. Learning features from enhanced function call graphs for Android malware detection. *Neurocomputing* 423 (2021), 301–307.

[17] Juan Cao, Junbo Guo, Xirong Li, Zhiwei Jin, Han Guo, and Jintao Li. 2018. Automatic Rumor Detection on Microblogs: A Survey. *CoRR* abs/1807.03505 (2018).

[18] Qiang Cao, Michael Sirivianos, Xiaowei Yang, and Tiago Pregueiro. 2012. Aiding the Detection of Fake Accounts in Large Scale Social Online Services. In *NSDI*. USENIX Association, 197–210.

[19] Qiang Cao, Xiaowei Yang, Jieqi Yu, and Christopher Palow. 2014. Uncovering Large Groups of Active Malicious Accounts in Online Social Networks. In *CCS*. ACM, 477–488.
[20] Carlos Castillo, Debora Donato, Aristides Gionis, Vanessa Murdock, and Fabrizio Silvestri. 2007. Know your neighbors: web spam detection using the web topology. In *SIGIR*. ACM, 423–430.
[21] Raghavendra Chalapathy and Sanjay Chawla. 2019. Deep learning for anomaly detection: A survey. *arXiv preprint arXiv:1901.03407* (2019).
[22] Jie Chen, Tengfei Ma, and Cao Xiao. 2018. FastGCN: Fast Learning with Graph Convolutional Networks via Importance Sampling. In *ICLR (Poster)*.
[23] Kai Chen, Peng Wang, Yeonjoon Lee, XiaoFeng Wang, Nan Zhang, Heqing Huang, Wei Zou, and Peng Liu. 2015. Finding Unknown Malice in 10 Seconds: Mass Vetting for New Threats at the Google-Play Scale. In *USENIX Security Symposium*. USENIX Association, 659–674.
[24] Ting Chen, Yuxiao Zhu, Zihao Li, Jiachi Chen, Xiaoqi Li, Xiapu Luo, Xiaodong Lin, and Xiaosong Zhang. 2018. Understanding Ethereum via Graph Analysis. In *INFOCOM*. IEEE, 1484–1492.
[25] Weili Chen, Jun Wu, Zibin Zheng, Chuan Chen, and Yuren Zhou. 2019. Market Manipulation of Bitcoin: Evidence from Mining the Mt. Gox Transaction Network. In *INFOCOM*. IEEE, 964–972.
[26] Zhengdao Chen, Lei Chen, Soledad Villar, and Joan Bruna. 2020. Can Graph Neural Networks Count Substructures?. In *NeurIPS*.
[27] Zhuo Chen, Na Lv, Kun Chen, Yanhui Zhang, and Weiting Gao. 2021. UAV network intrusion detection based on spatio-temporal graph convolutional network (in Chinese). *Journal of Beijing University of Aeronautics and Astronautics* 47, 05 (2021), 1068–1076.
[28] Flavio Chierichetti, Silvio Lattanzi, and Alessandro Panconesi. 2011. Rumor spreading in social networks. *Theor. Comput. Sci.* 412, 24 (2011), 2602–2610.
[29] Sudipta Chowdhury, Mojtaba Khanzadeh, Ravi Akula, Fangyan Zhang, Song Zhang, Hugh Medal, Mohammad Marufuzzaman, and Linkan Bian. 2017. Botnet detection using graph-based feature clustering. *Journal of Big Data* 4, 1 (2017), 1–23.
[30] Andrea Fronzetti Colladon and Elisa Remondi. 2017. Using social network analysis to prevent money laundering. *Expert Syst. Appl.* 67 (2017), 49–58.
[31] Jeffrey Dean and Sanjay Ghemawat. 2008. MapReduce: simplified data processing on large clusters. *Commun. ACM* 51, 1 (2008), 107–113.
[32] Sarthika Dhawan, Siva Charan Reddy Gangireddy, Shiv Kumar, and Tanmoy Chakraborty. 2019. Spotting Collective Behaviour of Online Frauds in Customer Reviews. In *IJCAI*. ijcai.org, 245–251.
[33] Hristo Djidjev, Gary Sandine, Curtis Storlie, and Scott Vander Wiel. 2011. Graph based statistical analysis of network traffic. In *Proceedings of the Ninth Workshop on Mining and Learning with Graphs*.
[34] Yingtong Dou, Zhiwei Liu, Li Sun, Yutong Deng, Hao Peng, and Philip S. Yu. 2020. Enhancing Graph Neural Network-based Fraud Detectors against Camouflaged Fraudsters. In *CIKM*. ACM, 315–324.
[35] Xu Duan, Jingzheng Wu, Tianyue Luo, Mutian Yang, and Yanjun Wu. 2010. Vulnerability mining method based on code property graph and attention BiLSTM (in Chinese). *Journal of Software* (2010).
[36] Scott W Duxbury and Dana L Haynie. 2018. The network structure of opioid distribution on a darknet cryptomarket. *Journal of quantitative criminology* 34, 4 (2018), 921–941.
[37] Ming Fan, Jun Liu, Xiapu Luo, Kai Chen, Tianyi Chen, Zhenzhou Tian, Xiaodong Zhang, Qinghua Zheng, and Ting Liu. 2016. Frequent Subgraph Based Familial Classification of Android Malware. In *ISSRE*. IEEE Computer Society, 24–35.
[38] Ming Fan, Jun Liu, Wei Wang, Haifei Li, Zhenzhou Tian, and Ting Liu. 2017. DAPASA: Detecting Android Piggybacked Apps Through Sensitive Subgraph Analysis. *IEEE Trans. Inf. Forensics Secur.* 12, 8 (2017), 1772–1785.
[39] Ming Fan, Xiapu Luo, Jun Liu, Meng Wang, Chunyin Nong, Qinghua Zheng, and Ting Liu. 2019. Graph embedding based familial analysis of Android malware using unsupervised learning. In *ICSE*. IEEE / ACM, 771–782.
[40] Yujie Fan, Shifu Hou, Yiming Zhang, Yanfang Ye, and Melih Abdulhayoglu. 2018. Gotcha-sly malware! scorpion a metagraph2vec based malware detection system. In *KDD*.
[41] Yujie Fan, Yanfang Ye, Qian Peng, Jianfei Zhang, Yiming Zhang, Xusheng Xiao, Chuan Shi, Qi Xiong, Fudong Shao, and Liang Zhao. 2020. Metagraph Aggregated Heterogeneous Graph Neural Network for Illicit Traded Product Identification in Underground Market. In *ICDM*.
[42] Qi Feng, Chengdong Feng, and Weijiang Hong. 2020. Graph Neural Network-based Vulnerability Predication. In *ICSME*. IEEE, 800–801.
[43] Qian Feng, Rundong Zhou, Chengcheng Xu, Yao Cheng, Brian Testa, and Heng Yin. 2016. Scalable Graph-based Bug Search for Firmware Images. In *CCS*. ACM, 480–491.
[44] Michael Fleder, Michael S. Kester, and Sudeep Pillai. 2015. Bitcoin Transaction Graph Analysis. *CoRR* abs/1502.01657 (2015).
[45] Anagi Gamachchi, Li Sun, and Serdar Boztas. 2018. A graph based framework for malicious insider threat detection. *arXiv preprint arXiv:1809.00141* (2018).
[46] Siva Charan Reddy Gangireddy, Deepak P, Cheng Long, and Tanmoy Chakraborty. 2020. Unsupervised Fake News Detection: A Graph-based Approach. In *HT*. ACM, 75–83.

[47] Sebastián García, Martin Grill, Jan Stiborek, and Alejandro Zunino. 2014. An empirical comparison of botnet detection methods. *Comput. Secur.* 45 (2014), 100–123.
[48] Vikas K. Garg, Stefanie Jegelka, and Tommi S. Jaakkola. 2020. Generalization and Representational Limits of Graph Neural Networks. In *ICML (Proceedings of Machine Learning Research, Vol. 119)*. PMLR, 3419–3430.
[49] Hugo Gascon, Fabian Yamaguchi, Daniel Arp, and Konrad Rieck. 2013. Structural detection of android malware using embedded call graphs. In *AISec*. ACM, 45–54.
[50] Justin Gilmer, Samuel S. Schoenholz, Patrick F. Riley, Oriol Vinyals, and George E. Dahl. 2017. Neural Message Passing for Quantum Chemistry. In *ICML (Proceedings of Machine Learning Research, Vol. 70)*. PMLR, 1263–1272.
[51] Luca Gioacchini, Luca Vassio, Marco Mellia, Idilio Drago, Zied Ben-Houidi, and Dario Rossi. 2021. DarkVec: automatic analysis of darknet traffic with word embeddings. In *CoNEXT*. ACM, 76–89.
[52] Palash Goyal and Emilio Ferrara. 2018. Graph embedding techniques, applications, and performance: A survey. *Knowl. Based Syst.* 151 (2018), 78–94.
[53] Kathrin Grosse, Nicolas Papernot, Praveen Manoharan, Michael Backes, and Patrick D. McDaniel. 2017. Adversarial Examples for Malware Detection. In *ESORICS (2) (Lecture Notes in Computer Science, Vol. 10493)*. Springer, 62–79.
[54] Aditya Grover and Jure Leskovec. 2016. node2vec: Scalable feature learning for networks. In *KDD*.
[55] Guofei Gu, Junjie Zhang, and Wenke Lee. 2008. BotSniffer: Detecting Botnet Command and Control Channels in Network Traffic. In *NDSS*. The Internet Society.
[56] Gisel Bastidas Guacho, Sara Abdali, Neil Shah, and Evangelos E. Papalexakis. 2018. Semi-supervised Content-Based Detection of Misinformation via Tensor Embeddings. In *ASONAM*. IEEE Computer Society, 322–325.
[57] Bin Guo, Yasan Ding, Yueheng Sun, Shuai Ma, Ke Li, and Zhiwen Yu. 2021. The mass, fake news, and cognition security. *Frontiers Comput. Sci.* 15, 3 (2021), 153806.
[58] Zoltán Gyöngyi, Hector Garcia-Molina, and Jan O. Pedersen. 2004. Combating Web Spam with TrustRank. In *VLDB*. Morgan Kaufmann, 576–587.
[59] William L. Hamilton, Rex Ying, and Jure Leskovec. 2017. Representation Learning on Graphs: Methods and Applications. *IEEE Data Eng. Bull.* 40, 3 (2017), 52–74.
[60] William L. Hamilton, Zhitao Ying, and Jure Leskovec. 2017. Inductive Representation Learning on Large Graphs. In *NIPS*. 1024–1034.
[61] Badis Hammi, Yacine Mohamed Idir, Sherali Zeadally, Rida Khatoun, and Jamel Nebhen. 2022. Is it Really Easy to Detect Sybil Attacks in C-ITS Environments: A Position Paper. *IEEE Trans. Intell. Transp. Syst.* 23, 10 (2022), 18273–18287.
[62] Robert Augustus Hardy and Julia R Norgaard. 2016. Reputation in the Internet black market: an empirical and theoretical analysis of the Deep Web. *Journal of Institutional Economics* 12, 3 (2016), 515–539.
[63] Chaoyang He, Keshav Balasubramanian, Emir Ceyani, Yu Rong, Peilin Zhao, Junzhou Huang, Murali Annavaram, and Salman Avestimehr. 2021. FedGraphNN: A Federated Learning System and Benchmark for Graph Neural Networks. *CoRR* abs/2104.07145 (2021).
[64] Wenxuan He, Gaopeng Gou, Cuicui Kang, Chang Liu, Zhen Li, and Gang Xiong. 2019. Malicious Domain Detection via Domain Relationship and Graph Models. In *IPCCC*. IEEE, 1–8.
[65] Atefeh Heydari, Mohammad ali Tavakoli, Naomie Salim, and Zahra Heydari. 2015. Detection of review spam: A survey. *Expert Syst. Appl.* 42, 7 (2015), 3634–3642.
[66] Bryan Hooi, Hyun Ah Song, Alex Beutel, Neil Shah, Kijung Shin, and Christos Faloutsos. 2016. FRAUDAR: Bounding Graph Fraud in the Face of Camouflage. In *KDD*. ACM, 895–904.
[67] Shifu Hou, Yujie Fan, Yiming Zhang, Yanfang Ye, Jingwei Lei, Wenqiang Wan, Jiabin Wang, Qi Xiong, and Fudong Shao. 2019. $\alpha$cyber: Enhancing robustness of android malware detection system against adversarial attacks on heterogeneous graph based model. In *CIKM*.
[68] Shifu Hou, Yanfang Ye, Yangqiu Song, and Melih Abdulhayoglu. 2017. Hindroid: An intelligent android malware detection system based on structured heterogeneous information network. In *KDD*.
[69] Binbin Hu, Zhiqiang Zhang, Chuan Shi, Jun Zhou, Xiaolong Li, and Yuan Qi. 2019. Cash-out user detection based on attributed heterogeneous information network with a hierarchical attention mechanism. In *AAAI*.
[70] Binbin Hu, Zhiqiang Zhang, Jun Zhou, Jingli Fang, Quanhui Jia, Yanming Fang, Quan Yu, and Yuan Qi. 2020. Loan Default Analysis with Multiplex Graph Learning. In *CIKM*.
[71] Xin Hu, Tzi-cker Chiueh, and Kang G Shin. 2009. Large-scale malware indexing using function-call graphs. In *CCS*.
[72] Jianjun Huang, Songming Han, Wei You, Wenchang Shi, Bin Liang, Jingzheng Wu, and Yanjun Wu. 2021. Hunting Vulnerable Smart Contracts via Graph Embedding Based Bytecode Matching. *IEEE Trans. Inf. Forensics Secur.* 16 (2021), 2144–2156.
[73] Luca Invernizzi, Stanislav Miskovic, Ruben Torres, Christopher Kruegel, Sabyasachi Saha, Giovanni Vigna, Sung-Ju Lee, and Marco Mellia. 2014. Nazca: Detecting Malware Distribution in Large-Scale Networks.. In *NDSS*.
[74] Sushil Jajodia, Paulo Shakarian, V. S. Subrahmanian, Vipin Swarup, and Cliff Wang (Eds.). 2015. *Cyber Warfare - Building the Scientific Foundation*. Advances in Information Security, Vol. 56. Springer.

[75] Tzung-Han Jeng, Yi-Ming Chen, Chien-Chih Chen, Chuan-Chiang Huang, and Kuo-Sen Chou. 2018. CC-Tracker: Interaction Profiling Bipartite Graph Mining for Malicious Network Activity Detection. In *DSC*. IEEE, 1–8.
[76] Yingsheng Ji, Zheng Zhang, Xinlei Tang, Jiachen Shen, Xi Zhang, and Guangwen Yang. 2022. Detecting Cash-out Users via Dense Subgraphs. In *KDD*. ACM, 687–697.
[77] Guanbo Jia, Paul Miller, Xin Hong, Harsha K. Kalutarage, and Tao Ban. 2019. Anomaly Detection in Network Traffic Using Dynamic Graph Mining with a Sparse Autoencoder. In *TrustCom/BigDataSE*. IEEE, 458–465.
[78] Yizhen Jia, Yinhao Xiao, Jiguo Yu, Xiuzhen Cheng, Zhenkai Liang, and Zhiguo Wan. 2018. A Novel Graph-based Mechanism for Identifying Traffic Vulnerabilitie in Smart Home IoT. In *INFOCOM*. IEEE, 1493–1501.
[79] Zhuoren Jiang, Zhe Gao, Yu Duan, Yangyang Kang, Changlong Sun, Qiong Zhang, and Xiaozhong Liu. 2020. Camouflaged Chinese Spam Content Detection with Semi-supervised Generative Active Learning. In *ACL*. Association for Computational Linguistics, 3080–3085.
[80] Zhuoren Jiang, Zhe Gao, Guoxiu He, Yangyang Kang, Changlong Sun, Qiong Zhang, Luo Si, and Xiaozhong Liu. 2019. Detect Camouflaged Spam Content via StoneSkipping: Graph and Text Joint Embedding for Chinese Character Variation Representation. In *EMNLP/IJCNLP (1)*. Association for Computational Linguistics, 6186–6195.
[81] Yu Jin, Nick Duffield, Patrick Haffner, Subhabrata Sen, and Z Zhang. 2011. Can't see forest through the trees? Understanding mixed network traffic graphs from application class distribution. In *Proc. of 9th Workshop on Mining and Learning with Graphs (MLG2011)*.
[82] Zhiwei Jin, Juan Cao, Yongdong Zhang, and Jiebo Luo. 2016. News Verification by Exploiting Conflicting Social Viewpoints in Microblogs. In *AAAI*. AAAI Press, 2972–2978.
[83] Parisa Kaghazgaran, James Caverlee, and Anna Cinzia Squicciarini. 2018. Combating Crowdsourced Review Manipulators: A Neighborhood-Based Approach. In *WSDM*. ACM, 306–314.
[84] Issa Khalil, Ting Yu, and Bei Guan. 2016. Discovering Malicious Domains through Passive DNS Data Graph Analysis. In *AsiaCCS*. ACM, 663–674.
[85] Abd Errahmane Kiouche, Sofiane Lagraa, Karima Amrouche, and Hamida Seba. 2021. A simple graph embedding for anomaly detection in a stream of heterogeneous labeled graphs. *Pattern Recognit.* 112 (2021), 107746.
[86] Thomas N. Kipf and Max Welling. 2017. Semi-Supervised Classification with Graph Convolutional Networks. In *ICLR (Poster)*.
[87] Danai Koutra, Tai-You Ke, U Kang, Duen Horng Chau, Hsing-Kuo Kenneth Pao, and Christos Faloutsos. 2011. Unifying Guilt-by-Association Approaches: Theorems and Fast Algorithms. In *ECML/PKDD (2) (Lecture Notes in Computer Science, Vol. 6912)*. Springer, 245–260.
[88] Srijan Kumar and Neil Shah. 2018. False Information on Web and Social Media: A Survey. *CoRR* abs/1804.08559 (2018).
[89] Bum Jun Kwon, Jayanta Mondal, Jiyong Jang, Leyla Bilge, and Tudor Dumitras. 2015. The Dropper Effect: Insights into Malware Distribution with Downloader Graph Analytics. In *CCS*. ACM, 1118–1129.
[90] Sofiane Lagraa, Jérôme François, Abdelkader Lahmadi, Marine Miner, Christian Hammerschmidt, and Radu State. 2017. BotGM: Unsupervised graph mining to detect botnets in traffic flows. In *2017 1st Cyber Security in Networking Conference (CSNet)*. IEEE, 1–8.
[91] Harjinder Singh Lallie, Lynsay A. Shepherd, Jason R. C. Nurse, Arnau Erola, Gregory Epiphaniou, Carsten Maple, and Xavier J. A. Bellekens. 2021. Cyber security in the age of COVID-19: A timeline and analysis of cyber-crime and cyber-attacks during the pandemic. *Comput. Secur.* 105 (2021), 102248.
[92] Do Quoc Le, Taeyoel Jeong, H. Eduardo Roman, and James Won-Ki Hong. 2011. Traffic dispersion graph based anomaly detection. In *SoICT*. ACM, 36–41.
[93] Kyumin Lee, Prithivi Tamilarasan, and James Caverlee. 2013. Crowdturfers, Campaigns, and Social Media: Tracking and Revealing Crowdsourced Manipulation of Social Media. In *ICWSM*. The AAAI Press.
[94] Kai Lei, Qiuai Fu, Jiake Ni, Feiyang Wang, Min Yang, and Kuai Xu. 2019. Detecting Malicious Domains with Behavioral Modeling and Graph Embedding. In *ICDCS*. IEEE, 601–611.
[95] Ao Li, Zhou Qin, Runshi Liu, Yiqun Yang, and Dong Li. 2019. Spam review detection with graph convolutional networks. In *CIKM*.
[96] Qian Li, Qingyuan Hu, Yong Qi, Saiyu Qi, Xinxing Liu, and Pengfei Gao. 2021. Semi-supervised two-phase familial analysis of Android malware with normalized graph embedding. *Knowl. Based Syst.* 218 (2021), 106802.
[97] Siyu Li, Jin Yang, Gang Liang, Tianrui Li, and Kui Zhao. 2022. SybilFlyover: Heterogeneous graph-based fake account detection model on social networks. *Knowl. Based Syst.* 258 (2022), 110038.
[98] Xiangfeng Li, Shenghua Liu, Zifeng Li, Xiaotian Han, Chuan Shi, Bryan Hooi, He Huang, and Xueqi Cheng. 2020. Flowscope: Spotting money laundering based on graphs. In *AAAI*.
[99] Yun Li, Chenlin Huang, Zhongfeng Wang, Lu Yuan, and Xiaochuan Wang. 2020. Survey of software vulnerability mining methods based on machine learning (in Chinese). *Journal of Software* 31, 7 (2020), 2040–2061.
[100] Yixuan Li, Oscar Martinez, Xing Chen, Yi Li, and John E. Hopcroft. 2016. In a World That Counts: Clustering and Detecting Fake Social Engagement at Scale. In *WWW*. ACM, 111–120.
[101] Chen Liang, ZQ LIU, Bin Liu, Jun Zhou, and Xiaolong Li. 2018. Who stole the postage? Fraud detection in return-freight insurance claims. In *KDD*.

[102] Chen Liang, Ziqi Liu, Bin Liu, Jun Zhou, Xiaolong Li, Shuang Yang, and Yuan Qi. 2019. Uncovering Insurance Fraud Conspiracy with Network Learning. In *SIGIR*. ACM, 1181–1184.
[103] Xiao Liang, Zheng Yang, Binghui Wang, Shaofeng Hu, Zijie Yang, Dong Yuan, Neil Zhenqiang Gong, Qi Li, and Fang He. 2021. Unveiling Fake Accounts at the Time of Registration: An Unsupervised Approach. In *KDD*. ACM, 3240–3250.
[104] Guanjun Lin, Sheng Wen, Qing-Long Han, Jun Zhang, and Yang Xiang. 2020. Software Vulnerability Detection Using Deep Neural Networks: A Survey. *Proc. IEEE* 108, 10 (2020), 1825–1848.
[105] Lin Liu, Wei-Tek Tsai, Md. Zakirul Alam Bhuiyan, Hao Peng, and Mingsheng Liu. 2022. Blockchain-enabled fraud discovery through abnormal smart contract detection on Ethereum. *Future Gener. Comput. Syst.* 128 (2022), 158–166.
[106] Ziqi Liu, Chaochao Chen, Longfei Li, Jun Zhou, Xiaolong Li, Le Song, and Yuan Qi. 2019. Geniepath: Graph neural networks with adaptive receptive paths. In *AAAI*.
[107] Ziqi Liu, Chaochao Chen, Xinxing Yang, Jun Zhou, Xiaolong Li, and Le Song. 2018. Heterogeneous graph neural networks for malicious account detection. In *CIKM*.
[108] Zhenguang Liu, Peng Qian, Xiang Wang, Lei Zhu, Qinming He, and Shouling Ji. 2021. Smart Contract Vulnerability Detection: From Pure Neural Network to Interpretable Graph Feature and Expert Pattern Fusion. In *IJCAI*. ijcai.org, 2751–2759.
[109] Zhenguang Liu, Peng Qian, Xiaoyang Wang, Yuan Zhuang, Lin Qiu, and Xun Wang. 2023. Combining Graph Neural Networks With Expert Knowledge for Smart Contract Vulnerability Detection. *IEEE Trans. Knowl. Data Eng.* 35, 2 (2023), 1296–1310.
[110] Yi-Ju Lu and Cheng-Te Li. 2020. GCAN: Graph-aware Co-Attention Networks for Explainable Fake News Detection on Social Media. In *ACL*. Association for Computational Linguistics, 505–514.
[111] Jun Ma, Danqing Zhang, Yun Wang, Yan Zhang, and Alexey Pozdnoukhov. 2018. GraphRAD: a graph-based risky account detection system. In *KDD*.
[112] Lingxiao Ma, Zhi Yang, Youshan Miao, Jilong Xue, Ming Wu, Lidong Zhou, and Yafei Dai. 2019. NeuGraph: Parallel Deep Neural Network Computation on Large Graphs. In *USENIX Annual Technical Conference*. USENIX Association, 443–458.
[113] Yao Ma and Jiliang Tang. 2021. *Deep Learning on Graphs*. Cambridge University Press.
[114] Samaneh Mahdavifar and Ali A Ghorbani. 2019. Application of deep learning to cybersecurity: A survey. *Neurocomputing* 347 (2019), 149–176.
[115] Tomas Mikolov, Kai Chen, Greg Corrado, and Jeffrey Dean. 2013. Efficient estimation of word representations in vector space. *arXiv preprint arXiv:1301.3781* (2013).
[116] Michael T. Mills and Nikolaos G. Bourbakis. 2014. Graph-Based Methods for Natural Language Processing and Understanding - A Survey and Analysis. *IEEE Trans. Syst. Man Cybern. Syst.* 44, 1 (2014), 59–71.
[117] Mohamed Nabeel, Issa M. Khalil, Bei Guan, and Ting Yu. 2020. Following Passive DNS Traces to Detect Stealthy Malicious Domains Via Graph Inference. *ACM Trans. Priv. Secur.* 23, 4 (2020), 17:1–17:36.
[118] Shishir Nagaraja, Prateek Mittal, Chi-Yao Hong, Matthew Caesar, and Nikita Borisov. 2010. BotGrep: Finding P2P Bots with Structured Graph Analysis.. In *USENIX security symposium*, Vol. 10. 95–110.
[119] Annamalai Narayanan, Mahinthan Chandramohan, Rajasekar Venkatesan, Lihui Chen, Yang Liu, and Shantanu Jaiswal. 2017. graph2vec: Learning Distributed Representations of Graphs. *CoRR* abs/1707.05005 (2017).
[120] Van-Hoang Nguyen, Kazunari Sugiyama, Preslav Nakov, and Min-Yen Kan. 2020. Fang: Leveraging social context for fake news detection using graph representation. In *CIKM*.
[121] Shirin Noekhah, Naomie Binti Salim, and Nor Hawaniah Zakaria. 2020. Opinion spam detection: Using multi-iterative graph-based model. *Inf. Process. Manag.* 57, 1 (2020).
[122] Anthony Palladino and Christopher J Thissen. 2018. Cyber anomaly detection using graph-node role-dynamics. *arXiv preprint arXiv:1812.02848* (2018).
[123] Shashank Pandit, Duen Horng Chau, Samuel Wang, and Christos Faloutsos. 2007. Netprobe: a fast and scalable system for fraud detection in online auction networks. In *WWW*. ACM, 201–210.
[124] Sergio Pastrana, Daniel R. Thomas, Alice Hutchings, and Richard Clayton. 2018. CrimeBB: Enabling Cybercrime Research on Underground Forums at Scale. In *WWW*. ACM, 1845–1854.
[125] Abdurrahman Pektaş and Tankut Acarman. 2020. Deep learning for effective Android malware detection using API call graph embeddings. *Soft Computing* 24, 2 (2020), 1027–1043.
[126] Bryan Perozzi, Rami Al-Rfou, and Steven Skiena. 2014. Deepwalk: Online learning of social representations. In *KDD*.
[127] Thai Pham and Steven Lee. 2016. Anomaly Detection in Bitcoin Network Using Unsupervised Learning Methods. *CoRR* abs/1611.03941 (2016).
[128] Tahereh Pourhabibi, Kok-Leong Ong, Booi Kam, and Yee Ling Boo. 2020. Fraud detection: A systematic literature review of graph-based anomaly detection approaches. *Decis. Support Syst.* 133 (2020), 113303.
[129] Brian A. Powell. 2020. Detecting malicious logins as graph anomalies. *J. Inf. Secur. Appl.* 54 (2020), 102557.
[130] Yanchen Qiao, Xiaochun Yun, and Yongzheng Zhang. 2016. How to Automatically Identify the Homology of Different Malware. In *Trustcom/BigDataSE/ISPA*. IEEE, 929–936.

[131] Meng Qu, Jian Tang, Jingbo Shang, Xiang Ren, Ming Zhang, and Jiawei Han. 2017. An attention-based collaboration framework for multi-view network representation learning. In *CIKM*.
[132] Stephen Ranshous, Cliff A. Joslyn, Sean Kreyling, Kathleen Nowak, Nagiza F. Samatova, Curtis L. West, and Samuel Winters. 2017. Exchange Pattern Mining in the Bitcoin Transaction Directed Hypergraph. In *Financial Cryptography Workshops (Lecture Notes in Computer Science, Vol. 10323)*. Springer, 248–263.
[133] Punit Rathore, Jayesh Soni, Nagarajan Prabakar, Marimuthu Palaniswami, and Paolo Santi. 2021. Identifying Groups of Fake Reviewers Using a Semisupervised Approach. *IEEE Trans. Comput. Soc. Syst.* 8, 6 (2021), 1369–1378.
[134] Jacob Ratkiewicz, Michael D. Conover, Mark R. Meiss, Bruno Gonçalves, Alessandro Flammini, and Filippo Menczer. 2011. Detecting and Tracking Political Abuse in Social Media. In *ICWSM*. The AAAI Press.
[135] Shebuti Rayana and Leman Akoglu. 2015. Collective opinion spam detection: Bridging review networks and metadata. In *KDD*.
[136] Xiaoxian Ren, Jie Chen, Chenyang Li, and Yixian Yang. 2018. Hazard assessment of IoT vulnerabilities correlation based on risk matrix (in Chinese). *Netinfo Security* 11 (2018).
[137] Shahbaz Rezaei and Xin Liu. 2019. Deep Learning for Encrypted Traffic Classification: An Overview. *IEEE Commun. Mag.* 57, 5 (2019), 76–81.
[138] Leonardo FR Ribeiro, Pedro HP Saverese, and Daniel R Figueiredo. 2017. struc2vec: Learning node representations from structural identity. In *KDD*.
[139] Sebastian Roschke, Feng Cheng, and Christoph Meinel. 2010. Using vulnerability information and attack graphs for intrusion detection. In *IAS*. IEEE, 68–73.
[140] Jay Rosenberg and Christiaan Beek. 2019. Examining Code Reuse Reveals Undiscovered Links among North Korea's Malware Families. *Examining Code Reuse Reveals Undiscovered Links among North Korea's Malware Families* (2019).
[141] Sherif Saad, Issa Traoré, Ali A. Ghorbani, Bassam Sayed, David Zhao, Wei Lu, John Felix, and Payman Hakimian. 2011. Detecting P2P botnets through network behavior analysis and machine learning. In *PST*. IEEE, 174–180.
[142] Hamidreza Sadreazami, Arash Mohammadi, Amir Asif, and Konstantinos N. Plataniotis. 2018. Distributed-Graph-Based Statistical Approach for Intrusion Detection in Cyber-Physical Systems. *IEEE Trans. Signal Inf. Process. over Networks* 4, 1 (2018), 137–147.
[143] Sagar Samtani, Hongyi Zhu, and Hsinchun Chen. 2020. Proactively Identifying Emerging Hacker Threats from the Dark Web: A Diachronic Graph Embedding Framework (D-GEF). *ACM Trans. Priv. Secur.* 23, 4 (2020), 21:1–21:33.
[144] Naw Safrin Sattar, Shaikh Arifuzzaman, Minhaz F. Zibran, and Md Mohiuddin Sakib. 2019. Detecting Web Spam in Webgraphs with Predictive Model Analysis. In *IEEE BigData*. IEEE, 4299–4308.
[145] Neil Shah, Alex Beutel, Brian Gallagher, and Christos Faloutsos. 2014. Spotting Suspicious Link Behavior with fBox: An Adversarial Perspective. In *ICDM*. IEEE Computer Society, 959–964.
[146] Neil Shah, Alex Beutel, Bryan Hooi, Leman Akoglu, Stephan Günnemann, Disha Makhija, Mohit Kumar, and Christos Faloutsos. 2016. EdgeCentric: Anomaly Detection in Edge-Attributed Networks. In *ICDM Workshops*. IEEE Computer Society, 327–334.
[147] Neil Shah, Hemank Lamba, Alex Beutel, and Christos Faloutsos. 2017. The Many Faces of Link Fraud. In *ICDM*. IEEE Computer Society, 1069–1074.
[148] Asim Shahzad, Nazri Mohd Nawi, Muhammad Zubair Rehman, and Abdullah Khan. 2021. An Improved Framework for Content- and Link-Based Web-Spam Detection: A Combined Approach. *Complex.* 2021 (2021), 6625739:1–6625739:18.
[149] Savva Shanaev, Arina Shuraeva, Mikhail Vasenin, and Maksim Kuznetsov. 2019. Cryptocurrency value and 51% attacks: evidence from event studies. *The Journal of Alternative Investments* 22, 3 (2019), 65–77.
[150] Yaoyao Shang, Shuangmao Yang, and Wei Wang. 2018. Botnet detection with hybrid analysis on flow based and graph based features of network traffic. In *International Conference on Cloud Computing and Security*. Springer, 612–621.
[151] Saeedreza Shehnepoor, Roberto Togneri, Wei Liu, and Mohammed Bennamoun. 2021. HIN-RNN: A Graph Representation Learning Neural Network for Fraudster Group Detection With No Handcrafted Features. *CoRR* abs/2105.11602 (2021).
[152] Jie Shen, Jiajun Zhou, Yunyi Xie, Shanqing Yu, and Qi Xuan. 2021. Identity Inference on Blockchain Using Graph Neural Network. In *BlockSys (Communications in Computer and Information Science, Vol. 1490)*. Springer, 3–17.
[153] Meng Shen, Jinpeng Zhang, Liehuang Zhu, Ke Xu, and Xiaojiang Du. 2021. Accurate Decentralized Application Identification via Encrypted Traffic Analysis Using Graph Neural Networks. *IEEE Trans. Inf. Forensics Secur.* 16 (2021), 2367–2380.
[154] Toshiki Shibahara, Yuta Takata, Mitsuaki Akiyama, Takeshi Yagi, and Takeshi Yada. 2017. Detecting Malicious Websites by Integrating Malicious, Benign, and Compromised Redirection Subgraph Similarities. In *COMPSAC (1)*. IEEE Computer Society, 655–664.
[155] Kai Shu, Deepak Mahudeswaran, Suhang Wang, and Huan Liu. 2020. Hierarchical Propagation Networks for Fake News Detection: Investigation and Exploitation. In *ICWSM*. AAAI Press, 626–637.
[156] Kai Shu, Suhang Wang, and Huan Liu. 2017. Exploiting Tri-Relationship for Fake News Detection. *CoRR* abs/1712.07709 (2017).
[157] Matteo Signorini, Wael Kanoun, and Roberto Di Pietro. 2018. ADvISE: Anomaly Detection tool for blockchaIn SystEms. In *SERVICES*. IEEE Computer Society, 65–66.
[158] Nikita Spirin and Jiawei Han. 2011. Survey on web spam detection: principles and algorithms. *SIGKDD Explor.* 13, 2 (2011), 50–64.

[159] Tatiana V. Stepanova and Dmitry P. Zegzhda. 2014. Applying Large-scale Adaptive Graphs to Modeling Internet of Things Security. In *SIN*. ACM, 479.
[160] Jack W. Stokes, Reid Andersen, Christian Seifert, and Kumar Chellapilla. 2010. WebCop: Locating Neighborhoods of Malware on the Web. In *LEET*. USENIX Association.
[161] W. Timothy Strayer, Robert Walsh, Carl Livadas, and David E. Lapsley. 2006. Detecting Botnets with Tight Command and Control. In *LCN*. IEEE Computer Society, 195–202.
[162] Gianluca Stringhini, Christopher Kruegel, and Giovanni Vigna. 2013. Shady paths: leveraging surfing crowds to detect malicious web pages. In *CCS*. ACM, 133–144.
[163] Boyu Sun, Wenyuan Yang, Mengqi Yan, Dehao Wu, Yuesheng Zhu, and Zhiqiang Bai. 2020. An Encrypted Traffic Classification Method Combining Graph Convolutional Network and Autoencoder. In *IPCCC*. IEEE, 1–8.
[164] Xiaoqing Sun, Zhiliang Wang, Jiahai Yang, and Xinran Liu. 2020. Deepdom: Malicious domain detection with scalable and heterogeneous graph convolutional networks. *Computers & Security* 99 (2020), 102057.
[165] Xiaoqing Sun, Jiahai Yang, Zhiliang Wang, and Heng Liu. 2020. HGDom: Heterogeneous Graph Convolutional Networks for Malicious Domain Detection. In *NOMS*. IEEE, 1–9.
[166] Yizhou Sun and Jiawei Han. 2012. Mining heterogeneous information networks: a structural analysis approach. *SIGKDD Explor.* 14, 2 (2012), 20–28.
[167] Yizhou Sun, Jiawei Han, Xifeng Yan, Philip S. Yu, and Tianyi Wu. 2011. PathSim: Meta Path-Based Top-K Similarity Search in Heterogeneous Information Networks. *Proc. VLDB Endow.* 4, 11 (2011), 992–1003.
[168] Toyotaro Suzumura, Yi Zhou, Natahalie Baracaldo, Guangnan Ye, Keith Houck, Ryo Kawahara, Ali Anwar, Lucia Larise Stavarache, Yuji Watanabe, Pablo Loyola, et al. 2019. Towards federated graph learning for collaborative financial crimes detection. *arXiv preprint arXiv:1909.12946* (2019).
[169] Mariarosaria Taddeo, Tom McCutcheon, and Luciano Floridi. 2019. Trusting artificial intelligence in cybersecurity is a double-edged sword. *Nat. Mach. Intell.* 1, 12 (2019), 557–560.
[170] Acar Tamersoy, Kevin Roundy, and Duen Horng Chau. 2014. Guilt by association: large scale malware detection by mining file-relation graphs. In *KDD*.
[171] Choon Lin Tan, Kang-Leng Chiew, Kelvin S. C. Yong, San-Nah Sze, Johari Abdullah, and Yakub Sebastian. 2020. A graph-theoretic approach for the detection of phishing webpages. *Comput. Secur.* 95 (2020), 101793.
[172] Jian Tang, Meng Qu, Mingzhe Wang, Ming Zhang, Jun Yan, and Qiaozhu Mei. 2015. Line: Large-scale information network embedding. In *WWW*.
[173] Lei Tang and Huan Liu. 2010. Graph mining applications to social network analysis. In *Managing and Mining Graph Data*. Springer, 487–513.
[174] Hau Tran, An Nguyen, Phuong Vo, and Tu Vu. 2017. DNS graph mining for malicious domain detection. In *IEEE BigData*. IEEE Computer Society, 4680–4685.
[175] Rudra M Tripathy, Amitabha Bagchi, and Sameep Mehta. 2010. A study of rumor control strategies on social networks. In *CIKM*.
[176] Alexey Tselykh, Margarita Knyazeva, Elena Popkova, Antonina Durfee, and Alexander Tselykh. 2016. An Attributed Graph Mining Approach to Detect Transfer Pricing Fraud. In *SIN*. ACM, 72–75.
[177] Petar Veličković, Guillem Cucurull, Arantxa Casanova, Adriana Romero, Pietro Lio, and Yoshua Bengio. 2017. Graph attention networks. *arXiv preprint arXiv:1710.10903* (2017).
[178] Andrei Venzhega, Polina Zhinalieva, and Nikolay Suboch. 2013. Graph-based malware distributors detection. In *WWW*.
[179] Rossouw von Solms and Johan Van Niekerk. 2013. From information security to cyber security. *Comput. Secur.* 38 (2013), 97–102.
[180] Daixin Wang, Jianbin Lin, Peng Cui, Quanhui Jia, Zhen Wang, Yanming Fang, Quan Yu, Jun Zhou, Shuang Yang, and Yuan Qi. 2019. A semi-supervised graph attentive network for financial fraud detection. In *ICDM*.
[181] Guan Wang, Sihong Xie, Bing Liu, and S Yu Philip. 2011. Review graph based online store review spammer detection. In *ICDM*.
[182] Guan Wang, Sihong Xie, Bing Liu, and Philip S. Yu. 2012. Identify Online Store Review Spammers via Social Review Graph. *ACM Trans. Intell. Syst. Technol.* 3, 4 (2012), 61:1–61:21.
[183] Huanting Wang, Guixin Ye, Zhanyong Tang, Shin Hwei Tan, Songfang Huang, Dingyi Fang, Yansong Feng, Lizhong Bian, and Zheng Wang. 2021. Combining Graph-Based Learning With Automated Data Collection for Code Vulnerability Detection. *IEEE Trans. Inf. Forensics Secur.* 16 (2021), 1943–1958.
[184] Jiaxin Wang, Yi Feng, Rui You, et al. 2019. Network security measurment based on dependency relationship graph and common vulnerability scoring system (in Chinese). *Journal of Computer Applications* 39, 6 (2019), 1719–1727.
[185] Jingjing Wang, Lansheng Han, Man Zhou, Wenkui Qian, and Dezhi An. 2021. Adaptive evaluation model of web spam based on link relation. *Trans. Emerg. Telecommun. Technol.* 32, 5 (2021).
[186] Jing Wang and Ioannis Ch Paschalidis. 2014. Botnet detection using social graph analysis. In *2014 52nd Annual Allerton Conference on Communication, Control, and Computing (Allerton)*. IEEE, 393–400.

[187] Jing Wang and Ioannis Ch Paschalidis. 2016. Botnet detection based on anomaly and community detection. *IEEE Transactions on Control of Network Systems* 4, 2 (2016), 392–404.
[188] Shen Wang, Zhengzhang Chen, Xiao Yu, Ding Li, Jingchao Ni, Lu-An Tang, Jiaping Gui, Zhichun Li, Haifeng Chen, and Philip S Yu. 2019. Heterogeneous Graph Matching Networks. *arXiv preprint arXiv:1910.08074* (2019).
[189] Shihan Wang and Takao Terano. 2015. Detecting rumor patterns in streaming social media. In *IEEE BigData*. IEEE Computer Society, 2709–2715.
[190] Wei Wang, Yaoyao Shang, Yongzhong He, Yidong Li, and Jiqiang Liu. 2020. BotMark: Automated botnet detection with hybrid analysis of flow-based and graph-based traffic behaviors. *Information Sciences* 511 (2020), 284–296.
[191] Xiao Wang, Deyu Bo, Chuan Shi, Shaohua Fan, Yanfang Ye, and Philip S. Yu. 2023. A Survey on Heterogeneous Graph Embedding: Methods, Techniques, Applications and Sources. *IEEE Trans. Big Data* 9, 2 (2023), 415–436.
[192] Xiao Wang, Houye Ji, Chuan Shi, Bai Wang, Yanfang Ye, Peng Cui, and Philip S Yu. 2019. Heterogeneous graph attention network. In *WWW*.
[193] Youze Wang, Shengsheng Qian, Jun Hu, Quan Fang, and Changsheng Xu. 2020. Fake News Detection via Knowledge-driven Multimodal Graph Convolutional Networks. In *ICMR*. ACM, 540–547.
[194] Yibo Wang and Wei Xu. 2018. Leveraging deep learning with LDA-based text analytics to detect automobile insurance fraud. *Decis. Support Syst.* 105 (2018), 87–95.
[195] Zhuo Wang, Songmin Gu, Xiangnan Zhao, and Xiaowei Xu. 2018. Graph-based review spammer group detection. *Knowl. Inf. Syst.* 55, 3 (2018), 571–597.
[196] Zhuo Wang, Tingting Hou, Dawei Song, Zhun Li, and Tianqi Kong. 2016. Detecting Review Spammer Groups via Bipartite Graph Projection. *Comput. J.* 59, 6 (2016), 861–874.
[197] Ziyu Wang, Nanqing Luo, and Pan Zhou. 2020. GuardHealth: Blockchain empowered secure data management and Graph Convolutional Network enabled anomaly detection in smart healthcare. *J. Parallel Distributed Comput.* 142 (2020), 1–12.
[198] Mark Weber, Jie Chen, Toyotaro Suzumura, Aldo Pareja, Tengfei Ma, Hiroki Kanezaki, Tim Kaler, Charles E Leiserson, and Tao B Schardl. 2018. Scalable graph learning for anti-money laundering: A first look. *arXiv preprint arXiv:1812.00076* (2018).
[199] Mark Weber, Giacomo Domeniconi, Jie Chen, Daniel Karl I Weidele, Claudio Bellei, Tom Robinson, and Charles E Leiserson. 2019. Anti-money laundering in bitcoin: Experimenting with graph convolutional networks for financial forensics. *arXiv preprint arXiv:1908.02591* (2019).
[200] Angus Wong and Alan Yeung. 2009. *Network infrastructure security*. Springer Science & Business Media.
[201] Jiajing Wu, Dan Lin, Zibin Zheng, and Qi Yuan. 2019. T-EDGE: Temporal WEighted MultiDiGraph Embedding for Ethereum Transaction Network Analysis. *CoRR* abs/1905.08038 (2019).
[202] Liang Wu and Huan Liu. 2018. Tracing Fake-News Footprints: Characterizing Social Media Messages by How They Propagate. In *WSDM*. ACM, 637–645.
[203] Zonghan Wu, Shirui Pan, Fengwen Chen, Guodong Long, Chengqi Zhang, and Philip S. Yu. 2021. A Comprehensive Survey on Graph Neural Networks. *IEEE Trans. Neural Networks Learn. Syst.* 32, 1 (2021), 4–24.
[204] Tobias Wüchner, Aleksander Cislak, Martín Ochoa, and Alexander Pretschner. 2019. Leveraging Compression-Based Graph Mining for Behavior-Based Malware Detection. *IEEE Trans. Dependable Secur. Comput.* 16, 1 (2019), 99–112.
[205] Cao Xiao, David Mandell Freeman, and Theodore Hwa. 2015. Detecting Clusters of Fake Accounts in Online Social Networks. In *AISec@CCS*. ACM, 91–101.
[206] Keyulu Xu, Weihua Hu, Jure Leskovec, and Stefanie Jegelka. 2019. How Powerful are Graph Neural Networks?. In *ICLR*.
[207] Xiaojun Xu, Chang Liu, Qian Feng, Heng Yin, Le Song, and Dawn Song. 2017. Neural Network-based Graph Embedding for Cross-Platform Binary Code Similarity Detection. In *CCS*. ACM, 363–376.
[208] Fabian Yamaguchi, Nico Golde, Daniel Arp, and Konrad Rieck. 2014. Modeling and Discovering Vulnerabilities with Code Property Graphs. In *IEEE Symposium on Security and Privacy*. IEEE Computer Society, 590–604.
[209] Cheng Yang, Zhiyuan Liu, Deli Zhao, Maosong Sun, and Edward Chang. 2015. Network representation learning with rich text information. In *IJCAI*.
[210] Chao Yang, Zhaoyan Xu, Guofei Gu, Vinod Yegneswaran, and Phillip A. Porras. 2014. DroidMiner: Automated Mining and Characterization of Fine-grained Malicious Behaviors in Android Applications. In *ESORICS (1) (Lecture Notes in Computer Science, Vol. 8712)*. Springer, 163–182.
[211] Shuo Yang, Kai Shu, Suhang Wang, Renjie Gu, Fan Wu, and Huan Liu. 2019. Unsupervised fake news detection on social media: A generative approach. In *AAAI*.
[212] Tianchi Yang, Linmei Hu, Chuan Shi, Houye Ji, Xiaoli Li, and Liqiang Nie. 2021. HGAT: Heterogeneous Graph Attention Networks for Semi-supervised Short Text Classification. *ACM Trans. Inf. Syst.* 39, 3 (2021), 32:1–32:29.
[213] Xiaoyu Yang, Yuefei Lyu, Tian Tian, Yifei Liu, Yudong Liu, and Xi Zhang. 2020. Rumor Detection on Social Media with Graph Structured Adversarial Learning.. In *IJCAI*.

[214] Yingguang Yang, Renyu Yang, Yangyang Li, Kai Cui, Zhiqin Yang, Yue Wang, Jie Xu, and Haiyong Xie. 2022. RoSGAS: Adaptive Social Bot Detection with Reinforced Self-Supervised GNN Architecture Search. *CoRR* abs/2206.06757 (2022).
[215] Junting Ye and Leman Akoglu. 2015. Discovering Opinion Spammer Groups by Network Footprints. In *ECML/PKDD (1) (Lecture Notes in Computer Science, Vol. 9284)*. Springer, 267–282.
[216] Jiexia Ye, Juanjuan Zhao, Kejiang Ye, and Chengzhong Xu. 2020. How to build a graph-based deep learning architecture in traffic domain: A survey. *IEEE Transactions on Intelligent Transportation Systems* (2020).
[217] Yanfang Ye, Lingwei Chen, Shifu Hou, William Hardy, and Xin Li. 2018. DeepAM: a heterogeneous deep learning framework for intelligent malware detection. *Knowl. Inf. Syst.* 54, 2 (2018), 265–285.
[218] Yanfang Ye, Shifu Hou, Lingwei Chen, Jingwei Lei, Wenqiang Wan, Jiabin Wang, Qi Xiong, and Fudong Shao. 2019. Out-of-sample node representation learning for heterogeneous graph in real-time android malware detection. In *IJCAI*.
[219] Yanfang Ye, Tao Li, Donald A. Adjeroh, and S. Sitharama Iyengar. 2017. A Survey on Malware Detection Using Data Mining Techniques. *ACM Comput. Surv.* 50, 3 (2017), 41:1–41:40.
[220] Yanfang Ye, Tao Li, Shenghuo Zhu, Weiwei Zhuang, Egemen Tas, Umesh Gupta, and Melih Abdulhayoglu. 2011. Combining file content and file relations for cloud based malware detection. In *KDD*.
[221] Hongji Yin and wei Chen. 2020. Server side request forgery vulnerability detection system with graph traversal algorithm (in Chinese). *Computer Engineering and Applications* 56, 19 (2020), 6.
[222] Zeping Yu, Rui Cao, Qiyi Tang, Sen Nie, Junzhou Huang, and Shi Wu. 2020. Order matters: Semantic-aware neural networks for binary code similarity detection. In *AAAI*.
[223] Chunyuan Yuan, Qianwen Ma, Wei Zhou, Jizhong Han, and Songlin Hu. 2019. Jointly embedding the local and global relations of heterogeneous graph for rumor detection. In *ICDM*.
[224] Dong Yuan, Yuanli Miao, Neil Zhenqiang Gong, Zheng Yang, Qi Li, Dawn Song, Qian Wang, and Xiao Liang. 2019. Detecting Fake Accounts in Online Social Networks at the Time of Registrations. In *CCS*. ACM, 1423–1438.
[225] Chuxu Zhang, Dongjin Song, Chao Huang, Ananthram Swami, and Nitesh V Chawla. 2019. Heterogeneous graph neural network. In *KDD*.
[226] Dalong Zhang, Xin Huang, Ziqi Liu, Jun Zhou, Zhiyang Hu, Xianzheng Song, Zhibang Ge, Lin Wang, Zhiqiang Zhang, and Yuan Qi. 2020. AGL: A Scalable System for Industrial-purpose Graph Machine Learning. *Proc. VLDB Endow.* 13, 12 (2020), 3125–3137.
[227] Daokun Zhang, Jie Yin, Xingquan Zhu, and Chengqi Zhang. 2018. MetaGraph2Vec: Complex Semantic Path Augmented Heterogeneous Network Embedding. In *PAKDD (2) (Lecture Notes in Computer Science, Vol. 10938)*. Springer, 196–208.
[228] Fuzhi Zhang, Xiaoyan Hao, Jinbo Chao, and Shuai Yuan. 2020. Label propagation-based approach for detecting review spammer groups on e-commerce websites. *Knowl. Based Syst.* 193 (2020), 105520.
[229] Huanding Zhang, Tao Shen, Fei Wu, Mingyang Yin, Hongxia Yang, and Chao Wu. 2021. Federated Graph Learning - A Position Paper. *CoRR* abs/2105.11099 (2021).
[230] Lei Zhang, Yong Cui, Jing Liu, Yong Jiang, and Jianping Wu. 2018. Application of machine learning in cyberspace security research (in Chinese). *Chinese Journal of Computers* 41, 9 (2018), 1943–1975.
[231] Lizhe Zhang, Zhaojun Gu, Bo He, and Shufa Liu. 2016. Multi-source attack pattern graph intrusion detection algorithm (in Chinese). *Computer Engineering and Design* 37, 11 (2016), 2909–2916.
[232] Shichang Zhang, Yozen Liu, Yizhou Sun, and Neil Shah. 2022. Graph-less Neural Networks: Teaching Old MLPs New Tricks Via Distillation. In *ICLR*.
[233] Yiming Zhang, Yujie Fan, Wei Song, Shifu Hou, Yanfang Ye, Xin Li, Liang Zhao, Chuan Shi, Jiabin Wang, and Qi Xiong. 2019. Your style your identity: Leveraging writing and photography styles for drug trafficker identification in darknet markets over attributed heterogeneous information network. In *WWW*.
[234] Yiming Zhang, Yujie Fan, Yanfang Ye, Liang Zhao, and Chuan Shi. 2019. Key player identification in underground forums over attributed heterogeneous information network embedding framework. In *CIKM*.
[235] Yiming Zhang, Yujie Fan, Yanfang Ye, Liang Zhao, Jiabin Wang, Qi Xiong, and Fudong Shao. 2018. KADetector: Automatic Identification of Key Actors in Online Hack Forums Based on Structured Heterogeneous Information Network. In *ICBK*. IEEE Computer Society, 154–161.
[236] Ya-Lin Zhang, Jun Zhou, Wenhao Zheng, Ji Feng, Longfei Li, Ziqi Liu, Ming Li, Zhiqiang Zhang, Chaochao Chen, Xiaolong Li, Yuan (Alan) Qi, and Zhi-Hua Zhou. 2019. Distributed Deep Forest and its Application to Automatic Detection of Cash-Out Fraud. *ACM Trans. Intell. Syst. Technol.* 10, 5 (2019), 55:1–55:19.
[237] Jun Zhao, Xudong Liu, Qiben Yan, Bo Li, Minglai Shao, and Hao Peng. 2020. Multi-attributed heterogeneous graph convolutional network for bot detection. *Information Sciences* 537 (2020), 380–393.
[238] Da Zheng, Chao Ma, Minjie Wang, Jinjing Zhou, Qidong Su, Xiang Song, Quan Gan, Zheng Zhang, and George Karypis. 2020. DistDGL: Distributed Graph Neural Network Training for Billion-Scale Graphs. In *IA3@SC*. IEEE, 36–44.
[239] Jingwei Zheng and Dagang Li. 2019. GCN-TC: Combining Trace Graph with Statistical Features for Network Traffic Classification. In *ICC*. IEEE, 1–6.

[240] Qiwei Zhong, Yang Liu, Xiang Ao, Binbin Hu, Jinghua Feng, Jiayu Tang, and Qing He. 2020. Financial defaulter detection on online credit payment via multi-view attributed heterogeneous information network. In *WWW*.
[241] Jie Zhou, Ganqu Cui, Shengding Hu, Zhengyan Zhang, Cheng Yang, Zhiyuan Liu, Lifeng Wang, Changcheng Li, and Maosong Sun. 2020. Graph neural networks: A review of methods and applications. *AI Open* 1 (2020), 57–81.
[242] Jiajun Zhou, Chenkai Hu, Shengbo Gong, Jiaying Xu, Jie Shen, and Qi Xuan. 2021. BlockGC: A Joint Learning Framework for Account Identity Inference on Blockchain with Graph Contrast. *CoRR* abs/2112.03659 (2021).
[243] Xinyi Zhou and Reza Zafarani. 2019. Network-based Fake News Detection: A Pattern-driven Approach. *SIGKDD Explor.* 21, 2 (2019), 48–60.
[244] Xinyi Zhou and Reza Zafarani. 2020. A Survey of Fake News: Fundamental Theories, Detection Methods, and Opportunities. *ACM Comput. Surv.* 53, 5 (2020), 109:1–109:40.
[245] Yaqin Zhou, Shangqing Liu, Jing Kai Siow, Xiaoning Du, and Yang Liu. 2019. Devign: Effective Vulnerability Identification by Learning Comprehensive Program Semantics via Graph Neural Networks. In *NeurIPS*. 10197–10207.
[246] Yuan Zhuang, Zhenguang Liu, Peng Qian, Qi Liu, Xiang Wang, and Qinming He. 2020. Smart Contract Vulnerability Detection using Graph Neural Network. In *IJCAI*. ijcai.org, 3283–3290.